\input pictex.tex   
\immediate\write10{Package DCpic 2002/05/16 v4.0}

\catcode`!=11 

\newcount\aux%
\newcount\auxa%
\newcount\auxb%
\newcount\m%
\newcount\n%
\newcount\x%
\newcount\y%
\newcount\xl%
\newcount\yl%
\newcount\d%
\newcount\dnm%
\newcount\xa%
\newcount\xb%
\newcount\xmed%
\newcount\xc%
\newcount\xd%
\newcount\ya%
\newcount\yb%
\newcount\ymed%
\newcount\yc%
\newcount\yd
\newcount\expansao%
\newcount\tipografo
\newcount\distanciaobjmor
\newcount\tipoarco
\newif\ifpara%
\newbox\caixa%
\newbox\caixaaux%
\newif\ifnvazia%
\newif\ifvazia%
\newif\ifcompara%
\newif\ifdiferentes%
\newcount\xaux%
\newcount\yaux%
\newcount\guardaauxa%
\newcount\alt%
\newcount\larg%
\newcount\prof%
\newcount\auxqx
\newcount\auxqy
\newif\ifajusta%
\newif\ifajustadist
\def\objPartida{}%
\def\objChegada{}%
\def\objNulo{}%


\def\!vazia{:}

\def\!pilhanvazia#1{\let\arg=#1%
\if:\arg\ \nvaziafalse\vaziatrue \else \nvaziatrue\vaziafalse\fi}

\def\!coloca#1#2{\edef\pilha{#1.#2}}

\def\!guarda(#1)(#2,#3)(#4,#5,#6){\def\id{#1}%
\xaux=#2%
\yaux=#3%
\alt=#4%
\larg=#5%
\prof=#6%
}

\def\!topaux#1.#2:{\!guarda#1}
\def\!topo#1{\expandafter\!topaux#1}

\def\!popaux#1.#2:{\def\pilha{#2:}}
\def\!retira#1{\expandafter\!popaux#1}

\def\!comparaaux#1#2{\let\argA=#1\let\argB=#2%
\ifx\argA\argB\comparatrue\diferentesfalse\else\comparafalse\diferentestrue\fi}

\def\!compara#1#2{\!comparaaux{#1}{#2}}

\def\!absoluto#1#2{\n=#1%
  \ifnum \n > 0
    #2=\n
  \else
    \multiply \n by -1
    #2=\n
  \fi}


\def\solidline{2}


\def\atleft{1}

\def\commdiag{0}


\def\!ajusta#1#2#3#4#5#6{\aux=#5%
  \let\auxobj=#6%
  \ifcase \tipografo    
    \ifnum\number\aux=10 
      \ajustadisttrue 
    \else
      \ajustadistfalse  
    \fi
  \else  
   \ajustadistfalse
  \fi
  \ifajustadist
   \let\pilhaaux=\pilha%
   \loop%
     \!topo{\pilha}%
     \!retira{\pilha}%
     \!compara{\id}{\auxobj}%
     \ifcompara\nvaziafalse \else\!pilhanvazia\pilha \fi%
     \ifnvazia%
   \repeat%
   \let\pilha=\pilhaaux%
   \ifvazia%
    \ifdiferentes%
     \larg=1310720
     \prof=655360%
     \alt=655360%
    \fi%
   \fi%
   \divide\larg by 131072
   \divide\prof by 65536
   \divide\alt by 65536
   \ifnum\number\y=\number\yl
    \advance\larg by 3
    \ifnum\number\larg>\aux
     #5=\larg
    \fi
   \else
    \ifnum\number\x=\number\xl
     \ifnum\number\yl>\number\y
      \ifnum\number\alt>\aux
       #5=\alt
      \fi
     \else
      \advance\prof by 5
      \ifnum\number\prof>\aux
       #5=\prof
      \fi
     \fi
    \else
     \auxqx=\x
     \advance\auxqx by -\xl
     \!absoluto{\auxqx}{\auxqx}%
     \auxqy=\y
     \advance\auxqy by -\yl
     \!absoluto{\auxqy}{\auxqy}%
     \ifnum\auxqx>\auxqy
      \ifnum\larg<10
       \larg=10
      \fi
      \advance\larg by 3
      #5=\larg
     \else
      \ifnum\yl>\y
       \ifnum\larg<10
        \larg=10
       \fi
      \advance\alt by 6
       #5=\alt
      \else
      \advance\prof by 11
       #5=\prof
      \fi
     \fi
    \fi
   \fi
\fi} 

\def\!raiz#1#2{\n=#1%
  \m=1%
  \loop
    \aux=\m%
    \advance \aux by 1%
    \multiply \aux by \aux%
    \ifnum \aux < \n%
      \advance \m by 1%
      \paratrue%
    \else\ifnum \aux=\n%
      \advance \m by 1%
      \paratrue%
       \else\parafalse%
       \fi
    \fi
  \ifpara%
  \repeat
#2=\m}

\def\!ucoord#1#2#3#4#5#6#7{\aux=#2%
  \advance \aux by -#1%
  \multiply \aux by #4%
  \divide \aux by #5%
  \ifnum #7 = -1 \multiply \aux by -1 \fi%
  \advance \aux by #3%
#6=\aux}

\def\!quadrado#1#2#3{\aux=#1%
  \advance \aux by -#2%
  \multiply \aux by \aux%
#3=\aux}

\def\!distnomemor#1#2#3#4#5#6{\setbox0=\hbox{#5}%
  \aux=#1
  \advance \aux by -#3
  \ifnum \aux=0
     \aux=\wd0 \divide \aux by 131072
     \advance \aux by 3
     #6=\aux
  \else
     \aux=#2
     \advance \aux by -#4
     \ifnum \aux=0
        \aux=\ht0 \advance \aux by \dp0 \divide \aux by 131072
        \advance \aux by 3
        #6=\aux%
     \else
     #6=3
     \fi
   \fi
}

\def\begindc#1{\!ifnextchar[{\!begindc{#1}}{\!begindc{#1}[30]}}
\def\!begindc#1[#2]{\beginpicture 
  \let\pilha=\!vazia
  \setcoordinatesystem units <1pt,1pt>
  \expansao=#2
  \ifcase #1
    \distanciaobjmor=10
    \tipoarco=0         
    \tipografo=0        
  \or
    \distanciaobjmor=2
    \tipoarco=0         
    \tipografo=1        
  \or
    \distanciaobjmor=1
    \tipoarco=2         
    \tipografo=2        
  \or
    \distanciaobjmor=8
    \tipoarco=0         
    \tipografo=3        
  \or
    \distanciaobjmor=8
    \tipoarco=2         
    \tipografo=4        
  \fi}

\def\enddc{\endpicture}

\def\mor{%
  \!ifnextchar({\!morxy}{\!morObjA}}
\def\!morxy(#1,#2){%
  \!ifnextchar({\!morxyl{#1}{#2}}{\!morObjB{#1}{#2}}}
\def\!morxyl#1#2(#3,#4){%
  \!ifnextchar[{\!mora{#1}{#2}{#3}{#4}}{\!mora{#1}{#2}{#3}{#4}[\number\distanciaobjmor,\number\distanciaobjmor]}}%
\def\!morObjA#1{%
 \let\pilhaaux=\pilha%
 \def\objPartida{#1}%
 \loop%
    \!topo\pilha%
    \!retira\pilha%
    \!compara{\id}{\objPartida}%
    \ifcompara \nvaziafalse \else \!pilhanvazia\pilha \fi%
   \ifnvazia%
 \repeat%
 \ifvazia%
  \ifdiferentes%
   Error: Incorrect label specification%
   \xaux=1%
   \yaux=1%
  \fi%
 \fi%
 \let\pilha=\pilhaaux%
 \!ifnextchar({\!morxyl{\number\xaux}{\number\yaux}}{\!morObjB{\number\xaux}{\number\yaux}}}
\def\!morObjB#1#2#3{%
  \x=#1
  \y=#2
 \def\objChegada{#3}%
 \let\pilhaaux=\pilha%
 \loop
    \!topo\pilha %
    \!retira\pilha%
    \!compara{\id}{\objChegada}%
    \ifcompara \nvaziafalse \else \!pilhanvazia\pilha \fi
   \ifnvazia
 \repeat
 \ifvazia
  \ifdiferentes%
   Error: Incorrect label specification
   \xaux=\x%
   \advance\xaux by \x%
   \yaux=\y%
   \advance\yaux by \y%
  \fi
 \fi
 \let\pilha=\pilhaaux
 \!ifnextchar[{\!mora{\number\x}{\number\y}{\number\xaux}{\number\yaux}}{\!mora{\number\x}{\number\y}{\number\xaux}{\number\yaux}[\number\distanciaobjmor,\number\distanciaobjmor]}}
\def\!mora#1#2#3#4[#5,#6]#7{%
  \!ifnextchar[{\!morb{#1}{#2}{#3}{#4}{#5}{#6}{#7}}{\!morb{#1}{#2}{#3}{#4}{#5}{#6}{#7}[1,\number\tipoarco] }}
\def\!morb#1#2#3#4#5#6#7[#8,#9]{\x=#1%
  \y=#2%
  \xl=#3%
  \yl=#4%
  \multiply \x by \expansao%
  \multiply \y by \expansao%
  \multiply \xl by \expansao%
  \multiply \yl by \expansao%
  \!quadrado{\number\x}{\number\xl}{\auxa}%
  \!quadrado{\number\y}{\number\yl}{\auxb}%
  \d=\auxa%
  \advance \d by \auxb%
  \!raiz{\d}{\d}%
  \auxa=#5
  \!compara{\objNulo}{\objPartida}%
  \ifdiferentes
   \!ajusta{\x}{\xl}{\y}{\yl}{\auxa}{\objPartida}%
   \ajustatrue
   \def\objPartida{}
  \fi
  \guardaauxa=\auxa
  \!ucoord{\number\x}{\number\xl}{\number\x}{\auxa}{\number\d}{\xa}{1}%
  \!ucoord{\number\y}{\number\yl}{\number\y}{\auxa}{\number\d}{\ya}{1}%
  \auxa=\d%
  \auxb=#6
  \!compara{\objNulo}{\objChegada}%
  \ifdiferentes
   \!ajusta{\x}{\xl}{\y}{\yl}{\auxb}{\objChegada}%
   \def\objChegada{}
  \fi
  \advance \auxa by -\auxb%
  \!ucoord{\number\x}{\number\xl}{\number\x}{\number\auxa}{\number\d}{\xb}{1}%
  \!ucoord{\number\y}{\number\yl}{\number\y}{\number\auxa}{\number\d}{\yb}{1}%
  \xmed=\xa%
  \advance \xmed by \xb%
  \divide \xmed by 2
  \ymed=\ya%
  \advance \ymed by \yb%
  \divide \ymed by 2
  \!distnomemor{\number\x}{\number\y}{\number\xl}{\number\yl}{#7}{\dnm}%
  \!ucoord{\number\y}{\number\yl}{\number\xmed}{\number\dnm}{\number\d}{\xc}{-#8}%
  \!ucoord{\number\x}{\number\xl}{\number\ymed}{\number\dnm}{\number\d}{\yc}{#8}%
\ifcase #9  
  \arrow <4pt> [.2,1.1] from {\xa} {\ya} to {\xb} {\yb}
\or  
  \setdashes
  \arrow <4pt> [.2,1.1] from {\xa} {\ya} to {\xb} {\yb}
  \setsolid
\or  
  \setlinear
  \plot {\xa} {\ya}  {\xb} {\yb} /
\or  
  \auxa=\guardaauxa
  \advance \auxa by 3%
 \!ucoord{\number\x}{\number\xl}{\number\x}{\number\auxa}{\number\d}{\xa}{1}%
 \!ucoord{\number\y}{\number\yl}{\number\y}{\number\auxa}{\number\d}{\ya}{1}%
 \!ucoord{\number\y}{\number\yl}{\number\xa}{3}{\number\d}{\xd}{-1}%
 \!ucoord{\number\x}{\number\xl}{\number\ya}{3}{\number\d}{\yd}{1}%
  \arrow <4pt> [.2,1.1] from {\xa} {\ya} to {\xb} {\yb}
  \circulararc -180 degrees from {\xa} {\ya} center at {\xd} {\yd}
\or  
  \auxa=3
 \!ucoord{\number\y}{\number\yl}{\number\xa}{\number\auxa}{\number\d}{\xmed}{-1}%
 \!ucoord{\number\x}{\number\xl}{\number\ya}{\number\auxa}{\number\d}{\ymed}{1}%
 \!ucoord{\number\y}{\number\yl}{\number\xa}{\number\auxa}{\number\d}{\xd}{1}%
 \!ucoord{\number\x}{\number\xl}{\number\ya}{\number\auxa}{\number\d}{\yd}{-1}%
  \arrow <4pt> [.2,1.1] from {\xa} {\ya} to {\xb} {\yb}
  \setlinear
  \plot {\xmed} {\ymed}  {\xd} {\yd} /
\fi
\auxa=\xl
\advance \auxa by -\x%
\ifnum \auxa=0 
  \put {#7} at {\xc} {\yc}
\else
  \auxb=\yl
  \advance \auxb by -\y%
  \ifnum \auxb=0 \put {#7} at {\xc} {\yc}
  \else 
    \ifnum \auxa > 0 
      \ifnum \auxb > 0
        \ifnum #8=1
          \put {#7} [rb] at {\xc} {\yc}
        \else 
          \put {#7} [lt] at {\xc} {\yc}
        \fi
      \else
        \ifnum #8=1
          \put {#7} [lb] at {\xc} {\yc}
        \else 
          \put {#7} [rt] at {\xc} {\yc}
        \fi
      \fi
    \else
      \ifnum \auxb > 0 
        \ifnum #8=1
          \put {#7} [rt] at {\xc} {\yc}
        \else 
          \put {#7} [lb] at {\xc} {\yc}
        \fi
      \else
        \ifnum #8=1
          \put {#7} [lt] at {\xc} {\yc}
        \else 
          \put {#7} [rb] at {\xc} {\yc}
        \fi
      \fi
    \fi
  \fi
\fi
}

\def\modifplot(#1{\!modifqcurve #1}
\def\!modifqcurve(#1,#2){\x=#1%
  \y=#2%
  \multiply \x by \expansao%
  \multiply \y by \expansao%
  \!start (\x,\y)
  \!modifQjoin}
\def\!modifQjoin(#1,#2)(#3,#4){\x=#1%
  \y=#2%
  \xl=#3%
  \yl=#4%
  \multiply \x by \expansao%
  \multiply \y by \expansao%
  \multiply \xl by \expansao%
  \multiply \yl by \expansao%
  \!qjoin (\x,\y) (\xl,\yl)             
  \!ifnextchar){\!fim}{\!modifQjoin}}
\def\!fim){\ignorespaces}

\def\setaxy(#1{\!pontosxy #1}
\def\!pontosxy(#1,#2){%
  \!maispontosxy}
\def\!maispontosxy(#1,#2)(#3,#4){%
  \!ifnextchar){\!fimxy#3,#4}{\!maispontosxy}}
\def\!fimxy#1,#2){\x=#1%
  \y=#2
  \multiply \x by \expansao
  \multiply \y by \expansao
  \xl=\x%
  \yl=\y%
  \aux=1%
  \multiply \aux by \auxa%
  \advance\xl by \aux%
  \aux=1%
  \multiply \aux by \auxb%
  \advance\yl by \aux%
  \arrow <4pt> [.2,1.1] from {\x} {\y} to {\xl} {\yl}}

\def\cmor#1 #2(#3,#4)#5{%
  \!ifnextchar[{\!cmora{#1}{#2}{#3}{#4}{#5}}{\!cmora{#1}{#2}{#3}{#4}{#5}[0] }}
\def\!cmora#1#2#3#4#5[#6]{%
  \ifcase #2
      \auxa=0
      \auxb=1
    \or
      \auxa=0
      \auxb=-1
    \or
      \auxa=1
      \auxb=0
    \or
      \auxa=-1
      \auxb=0
    \fi  
  \ifcase #6  
    \modifplot#1
    \setaxy#1
  \or  
    \setdashes
    \modifplot#1
    \setaxy#1
    \setsolid
  \or  
    \modifplot#1
  \fi  
  \x=#3%
  \y=#4%
  \multiply \x by \expansao%
  \multiply \y by \expansao%
  \put {#5} at {\x} {\y}}

\def\obj(#1,#2){%
  \!ifnextchar[{\!obja{#1}{#2}}{\!obja{#1}{#2}[Nulo]}}
\def\!obja#1#2[#3]#4{%
  \!ifnextchar[{\!objb{#1}{#2}{#3}{#4}}{\!objb{#1}{#2}{#3}{#4}[1]}}
\def\!objb#1#2#3#4[#5]{%
  \x=#1%
  \y=#2%
  \def\!pinta{\normalsize$\bullet$}
  \def\!nulo{Nulo}%
  \def\!arg{#3}%
  \!compara{\!arg}{\!nulo}%
  \ifcompara\def\!arg{#4}\fi%
  \multiply \x by \expansao%
  \multiply \y by \expansao%
  \setbox\caixa=\hbox{#4}%
  \!coloca{(\!arg)(#1,#2)(\number\ht\caixa,\number\wd\caixa,\number\dp\caixa)}{\pilha}%
  \auxa=\wd\caixa \divide \auxa by 131072 
  \advance \auxa by 5
  \auxb=\ht\caixa
  \advance \auxb by \number\dp\caixa
  \divide \auxb by 131072 
  \advance \auxb by 5
  \ifcase \tipografo    
    \put{#4} at {\x} {\y}
  \or                   
    \ifcase #5 
      \put{#4} at {\x} {\y}
    \or        
      \put{\!pinta} at {\x} {\y}
      \advance \y by \number\auxb  
      \put{#4} at {\x} {\y}
    \or        
      \put{\!pinta} at {\x} {\y}
      \advance \auxa by -2  
      \advance \auxb by -2  
      \advance \x by \number\auxa  
      \advance \y by \number\auxb  
      \put{#4} at {\x} {\y}   
    \or        
      \put{\!pinta} at {\x} {\y}
      \advance \x by \number\auxa  
      \put{#4} at {\x} {\y}   
    \or        
      \put{\!pinta} at {\x} {\y}
      \advance \auxa by -2  
      \advance \auxb by -2  
      \advance \x by \number\auxa  
      \advance \y by -\number\auxb  
      \put{#4} at {\x} {\y}   
    \or        
      \put{\!pinta} at {\x} {\y}
      \advance \y by -\number\auxb  
      \put{#4} at {\x} {\y}   
    \or        
      \put{\!pinta} at {\x} {\y}
      \advance \auxa by -2  
      \advance \auxb by -2  
      \advance \x by -\number\auxa  
      \advance \y by -\number\auxb  
      \put{#4} at {\x} {\y}   
    \or        
      \put{\!pinta} at {\x} {\y}
      \advance \x by -\number\auxa  
      \put{#4} at {\x} {\y}   
    \or        
      \put{\!pinta} at {\x} {\y}
      \advance \auxa by -2  
      \advance \auxb by -2  
      \advance \x by -\number\auxa  
      \advance \y by \number\auxb  
      \put{#4} at {\x} {\y}   
    \fi
  \or                   
    \ifcase #5 
      \put{#4} at {\x} {\y}
    \or        
      \put{\!pinta} at {\x} {\y}
      \advance \y by \number\auxb  
      \put{#4} at {\x} {\y}
    \or        
      \put{\!pinta} at {\x} {\y}
      \advance \auxa by -2  
      \advance \auxb by -2  
      \advance \x by \number\auxa  
      \advance \y by \number\auxb  
      \put{#4} at {\x} {\y}   
    \or        
      \put{\!pinta} at {\x} {\y}
      \advance \x by \number\auxa  
      \put{#4} at {\x} {\y}   
    \or        
      \put{\!pinta} at {\x} {\y}
      \advance \auxa by -2  
      \advance \auxb by -2
      \advance \x by \number\auxa  
      \advance \y by -\number\auxb 
      \put{#4} at {\x} {\y}   
    \or        
      \put{\!pinta} at {\x} {\y}
      \advance \y by -\number\auxb 
      \put{#4} at {\x} {\y}   
    \or        
      \put{\!pinta} at {\x} {\y}
      \advance \auxa by -2  
      \advance \auxb by -2
      \advance \x by -\number\auxa 
      \advance \y by -\number\auxb 
      \put{#4} at {\x} {\y}   
    \or        
      \put{\!pinta} at {\x} {\y}
      \advance \x by -\number\auxa 
      \put{#4} at {\x} {\y}   
    \or        
      \put{\!pinta} at {\x} {\y}
      \advance \auxa by -2  
      \advance \auxb by -2
      \advance \x by -\number\auxa 
      \advance \y by \number\auxb  
      \put{#4} at {\x} {\y}   
    \fi
   \else 
     \ifnum\auxa<\auxb 
       \aux=\auxb
     \else
       \aux=\auxa
     \fi
     \ifdim\wd\caixa<1em
       \dimen99 = 1em
       \aux=\dimen99 \divide \aux by 131072 
       \advance \aux by 5
     \fi
     \advance\aux by -2 
     \multiply\aux by 2 %
     \ifnum\aux<30
       \put{\circle{\aux}} [Bl] at {\x} {\y}
     \else
       \multiply\auxa by 2
       \multiply\auxb by 2
       \put{\oval(\auxa,\auxb)} [Bl] at {\x} {\y}
     \fi
     \put{#4} at {\x} {\y}
   \fi   
}

\catcode`!=12 

  \input miniltx
  \def\Gin@driver{pdftex.def}
  \input color.sty
  \input graphicx.sty
  \resetatcatcode

%
%

%
%
%
%

\def\Serif{cmr}
\def\SerifBold{cmbx}
\def\SerifItalics{cmti}
\def\SerifSlanted{cmsl}
\def\SerifBoldItalics{cmbxti}
\def\SansSerif{cmss}
\def\SansSerifBold{cmssbx}
\def\SansSerifItalics{cmssi}
\def\SansSerifSlanted{cmssi}
\def\Math{cmmi}
\def\Symbols{cmsy}
\def\MathBold{cmmib}
\def\MoreSymbols{cmex}
\def\Typewriter{cmtt}
\def\Gothic{eufm}
\def\Double{msbm}
\def\Relazioni{msam}

= 			\Serif10 			at 5pt
= 		\SerifBold10 		at 5pt
= 	\SerifItalics10 	at 5pt
=		\SerifSlanted10 	at 5pt
=	\SerifBoldItalics10	at 5pt
= 		\SansSerif10 		at 5pt
=	\SansSerifBold10	at 5pt
=	\SansSerifItalics10	at 5pt
=	\SansSerifSlanted10	at 5pt
=				\Math10				at 5pt
=			\MathBold10			at 5pt
=			\Symbols10			at 5pt
=		\MoreSymbols10		at 5pt
=		\Typewriter10		at 5pt
=			\Gothic10			at 5pt
=			\Double10			at 5pt

= 			\Serif10 			at 7pt
= 		\SerifBold10 		at 7pt
= 	\SerifItalics10 	at 7pt
=	\SerifSlanted10 	at 7pt
=\SerifBoldItalics10	at 7pt
= 		\SansSerif10 		at 7pt
= 	\SansSerifBold10 	at 7pt
=\SansSerifItalics10	at 7pt
=\SansSerifSlanted10	at 7pt
=			\Math10				at 7pt
=		\MathBold10			at 7pt
=			\Symbols10			at 7pt
=		\MoreSymbols10		at 7pt
=		\Typewriter10		at 7pt
=			\Gothic10			at 7pt
=			\Double10			at 7pt

= 			\Serif10 			at 8pt
= 		\SerifBold10 		at 8pt
= 	\SerifItalics10 	at 8pt
=	\SerifSlanted10 	at 8pt
=\SerifBoldItalics10	at 8pt
= 		\SansSerif10 		at 8pt
= 	\SansSerifBold10 	at 8pt
=\SansSerifItalics10 at 8pt
=\SansSerifSlanted10 at 8pt
=			\Math10				at 8pt
=		\MathBold10			at 8pt
=			\Symbols10			at 8pt
=		\MoreSymbols10		at 8pt
=		\Typewriter10		at 8pt
=			\Gothic10			at 8pt
=			\Double10			at 8pt

= 			\Serif10 			at 10pt
= 		\SerifBold10 		at 10pt
= 		\SerifItalics10 	at 10pt
=		\SerifSlanted10 	at 10pt
=	\SerifBoldItalics10	at 10pt
= 		\SansSerif10 		at 10pt
= 	\SansSerifBold10 	at 10pt
= 	\SansSerifItalics10 at 10pt
= 	\SansSerifSlanted10 at 10pt
=				\Math10				at 10pt
=			\MathBold10			at 10pt
=			\Symbols10			at 10pt
=		\MoreSymbols10		at 10pt
=		\Typewriter10		at 10pt
=			\Gothic10			at 10pt
=			\Double10			at 10pt
=			\Relazioni10			at 10pt

= 				\Serif10 			at 12pt
= 			\SerifBold10 		at 12pt
= 		\SerifItalics10 	at 12pt
=		\SerifSlanted10 	at 12pt
=	\SerifBoldItalics10	at 12pt
= 			\SansSerif10 		at 12pt
= 		\SansSerifBold10 	at 12pt
= 	\SansSerifItalics10 at 12pt
= 	\SansSerifSlanted10 at 12pt
=				\Math10				at 12pt
=			\MathBold10			at 12pt
=			\Symbols10			at 12pt
=		\MoreSymbols10		at 12pt
=			\Typewriter10		at 12pt
=				\Gothic10			at 12pt
=				\Double10			at 12pt

= 			\Serif10 			at 14pt
= 		\SerifBold10 		at 14pt
= 	\SerifItalics10 	at 14pt
=		\SerifSlanted10 	at 14pt
=	\SerifBoldItalics10	at 14pt
= 		\SansSerif10 		at 14pt
= 	\SansSerifBold10 	at 14pt
= \SansSerifSlanted10 at 14pt
= \SansSerifItalics10 at 14pt
=				\Math10				at 14pt
=			\MathBold10			at 14pt
=			\Symbols10			at 14pt
=		\MoreSymbols10		at 14pt
=		\Typewriter10		at 14pt
=			\Gothic10			at 14pt
=			\Double10			at 14pt

\def\NormalStyle{\parindent=5pt\parskip=3pt\normalbaselineskip=14pt%
\def\nt{\tenSerif}%
\def\rm{\fam0\tenSerif}%
\textfont0=\tenSerif\scriptfont0=\sevenSerif\scriptscriptfont0=\fiveSerif
\textfont1=\tenMath\scriptfont1=\sevenMath\scriptscriptfont1=\fiveMath
\textfont2=\tenSymbols\scriptfont2=\sevenSymbols\scriptscriptfont2=\fiveSymbols
\textfont3=\tenMoreSymbols\scriptfont3=\sevenMoreSymbols\scriptscriptfont3=\fiveMoreSymbols
\textfont\itfam=\tenSerifItalics\def\it{\fam\itfam\tenSerifItalics}%
\textfont\slfam=\tenSerifSlanted\def\sl{\fam\slfam\tenSerifSlanted}%
\textfont\ttfam=\tenTypewriter\def\tt{\fam\ttfam\tenTypewriter}%
\textfont\bffam=\tenSerifBold%
\def\bf{\fam\bffam\tenSerifBold}\scriptfont\bffam=\sevenSerifBold\scriptscriptfont\bffam=\fiveSerifBold%
\def\cal{\tenSymbols}%
\def\greekbold{\tenMathBold}%
\def\gothic{\tenGothic}%
\def\Bbb{\tenDouble}%
\def\LieFont{\tenSerifItalics}%
\nt\normalbaselines\baselineskip=14pt%
}

\def\TitleStyle{\parindent=0pt\parskip=0pt\normalbaselineskip=15pt%
\def\nt{\fourteenSansSerifBold}%
\def\rm{\fam0\fourteenSansSerifBold}%
\textfont0=\fourteenSansSerifBold\scriptfont0=\tenSansSerifBold\scriptscriptfont0=\eightSansSerifBold
\textfont1=\fourteenMath\scriptfont1=\tenMath\scriptscriptfont1=\eightMath
\textfont2=\fourteenSymbols\scriptfont2=\tenSymbols\scriptscriptfont2=\eightSymbols
\textfont3=\fourteenMoreSymbols\scriptfont3=\tenMoreSymbols\scriptscriptfont3=\eightMoreSymbols
\textfont\itfam=\fourteenSansSerifItalics\def\it{\fam\itfam\fourteenSansSerifItalics}%
\textfont\slfam=\fourteenSansSerifSlanted\def\sl{\fam\slfam\fourteenSerifSansSlanted}%
\textfont\ttfam=\fourteenTypewriter\def\tt{\fam\ttfam\fourteenTypewriter}%
\textfont\bffam=\fourteenSansSerif%
\def\bf{\fam\bffam\fourteenSansSerif}\scriptfont\bffam=\tenSansSerif\scriptscriptfont\bffam=\eightSansSerif%
\def\cal{\fourteenSymbols}%
\def\greekbold{\fourteenMathBold}%
\def\gothic{\fourteenGothic}%
\def\Bbb{\fourteenDouble}%
\def\LieFont{\fourteenSerifItalics}%
\nt\normalbaselines\baselineskip=15pt%
}

\def\PartStyle{\parindent=0pt\parskip=0pt\normalbaselineskip=15pt%
\def\nt{\fourteenSansSerifBold}%
\def\rm{\fam0\fourteenSansSerifBold}%
\textfont0=\fourteenSansSerifBold\scriptfont0=\tenSansSerifBold\scriptscriptfont0=\eightSansSerifBold
\textfont1=\fourteenMath\scriptfont1=\tenMath\scriptscriptfont1=\eightMath
\textfont2=\fourteenSymbols\scriptfont2=\tenSymbols\scriptscriptfont2=\eightSymbols
\textfont3=\fourteenMoreSymbols\scriptfont3=\tenMoreSymbols\scriptscriptfont3=\eightMoreSymbols
\textfont\itfam=\fourteenSansSerifItalics\def\it{\fam\itfam\fourteenSansSerifItalics}%
\textfont\slfam=\fourteenSansSerifSlanted\def\sl{\fam\slfam\fourteenSerifSansSlanted}%
\textfont\ttfam=\fourteenTypewriter\def\tt{\fam\ttfam\fourteenTypewriter}%
\textfont\bffam=\fourteenSansSerif%
\def\bf{\fam\bffam\fourteenSansSerif}\scriptfont\bffam=\tenSansSerif\scriptscriptfont\bffam=\eightSansSerif%
\def\cal{\fourteenSymbols}%
\def\greekbold{\fourteenMathBold}%
\def\gothic{\fourteenGothic}%
\def\Bbb{\fourteenDouble}%
\def\LieFont{\fourteenSerifItalics}%
\nt\normalbaselines\baselineskip=15pt%
}

\def\ChapterStyle{\parindent=0pt\parskip=0pt\normalbaselineskip=15pt%
\def\nt{\fourteenSansSerifBold}%
\def\rm{\fam0\fourteenSansSerifBold}%
\textfont0=\fourteenSansSerifBold\scriptfont0=\tenSansSerifBold\scriptscriptfont0=\eightSansSerifBold
\textfont1=\fourteenMath\scriptfont1=\tenMath\scriptscriptfont1=\eightMath
\textfont2=\fourteenSymbols\scriptfont2=\tenSymbols\scriptscriptfont2=\eightSymbols
\textfont3=\fourteenMoreSymbols\scriptfont3=\tenMoreSymbols\scriptscriptfont3=\eightMoreSymbols
\textfont\itfam=\fourteenSansSerifItalics\def\it{\fam\itfam\fourteenSansSerifItalics}%
\textfont\slfam=\fourteenSansSerifSlanted\def\sl{\fam\slfam\fourteenSerifSansSlanted}%
\textfont\ttfam=\fourteenTypewriter\def\tt{\fam\ttfam\fourteenTypewriter}%
\textfont\bffam=\fourteenSansSerif%
\def\bf{\fam\bffam\fourteenSansSerif}\scriptfont\bffam=\tenSansSerif\scriptscriptfont\bffam=\eightSansSerif%
\def\cal{\fourteenSymbols}%
\def\greekbold{\fourteenMathBold}%
\def\gothic{\fourteenGothic}%
\def\Bbb{\fourteenDouble}%
\def\LieFont{\fourteenSerifItalics}%
\nt\normalbaselines\baselineskip=15pt%
}

\def\SectionStyle{\parindent=0pt\parskip=0pt\normalbaselineskip=13pt%
\def\nt{\twelveSansSerifBold}%
\def\rm{\fam0\twelveSansSerifBold}%
\textfont0=\twelveSansSerifBold\scriptfont0=\eightSansSerifBold\scriptscriptfont0=\eightSansSerifBold
\textfont1=\twelveMath\scriptfont1=\eightMath\scriptscriptfont1=\eightMath
\textfont2=\twelveSymbols\scriptfont2=\eightSymbols\scriptscriptfont2=\eightSymbols
\textfont3=\twelveMoreSymbols\scriptfont3=\eightMoreSymbols\scriptscriptfont3=\eightMoreSymbols
\textfont\itfam=\twelveSansSerifItalics\def\it{\fam\itfam\twelveSansSerifItalics}%
\textfont\slfam=\twelveSansSerifSlanted\def\sl{\fam\slfam\twelveSerifSansSlanted}%
\textfont\ttfam=\twelveTypewriter\def\tt{\fam\ttfam\twelveTypewriter}%
\textfont\bffam=\twelveSansSerif%
\def\bf{\fam\bffam\twelveSansSerif}\scriptfont\bffam=\eightSansSerif\scriptscriptfont\bffam=\eightSansSerif%
\def\cal{\twelveSymbols}%
\def\bg{\twelveMathBold}%
\def\gothic{\twelveGothic}%
\def\Bbb{\twelveDouble}%
\def\LieFont{\twelveSerifItalics}%
\nt\normalbaselines\baselineskip=13pt%
}

\def\SubSectionStyle{\parindent=0pt\parskip=0pt\normalbaselineskip=13pt%
\def\nt{\twelveSansSerifItalics}%
\def\rm{\fam0\twelveSansSerifItalics}%
\textfont0=\twelveSansSerifItalics\scriptfont0=\eightSansSerifItalics\scriptscriptfont0=\eightSansSerifItalics%
\textfont1=\twelveMath\scriptfont1=\eightMath\scriptscriptfont1=\eightMath%
\textfont2=\twelveSymbols\scriptfont2=\eightSymbols\scriptscriptfont2=\eightSymbols%
\textfont3=\twelveMoreSymbols\scriptfont3=\eightMoreSymbols\scriptscriptfont3=\eightMoreSymbols%
\textfont\itfam=\twelveSansSerif\def\it{\fam\itfam\twelveSansSerif}%
\textfont\slfam=\twelveSansSerifSlanted\def\sl{\fam\slfam\twelveSerifSansSlanted}%
\textfont\ttfam=\twelveTypewriter\def\tt{\fam\ttfam\twelveTypewriter}%
\textfont\bffam=\twelveSansSerifBold%
\def\bf{\fam\bffam\twelveSansSerifBold}\scriptfont\bffam=\eightSansSerifBold\scriptscriptfont\bffam=\eightSansSerifBold%
\def\cal{\twelveSymbols}%
\def\greekbold{\twelveMathBold}%
\def\gothic{\twelveGothic}%
\def\Bbb{\twelveDouble}%
\def\LieFont{\twelveSerifItalics}%
\nt\normalbaselines\baselineskip=13pt%
}

\def\AuthorStyle{\parindent=0pt\parskip=0pt\normalbaselineskip=14pt%
\def\nt{\tenSerif}%
\def\rm{\fam0\tenSerif}%
\textfont0=\tenSerif\scriptfont0=\sevenSerif\scriptscriptfont0=\fiveSerif
\textfont1=\tenMath\scriptfont1=\sevenMath\scriptscriptfont1=\fiveMath
\textfont2=\tenSymbols\scriptfont2=\sevenSymbols\scriptscriptfont2=\fiveSymbols
\textfont3=\tenMoreSymbols\scriptfont3=\sevenMoreSymbols\scriptscriptfont3=\fiveMoreSymbols
\textfont\itfam=\tenSerifItalics\def\it{\fam\itfam\tenSerifItalics}%
\textfont\slfam=\tenSerifSlanted\def\sl{\fam\slfam\tenSerifSlanted}%
\textfont\ttfam=\tenTypewriter\def\tt{\fam\ttfam\tenTypewriter}%
\textfont\bffam=\tenSerifBold%
\def\bf{\fam\bffam\tenSerifBold}\scriptfont\bffam=\sevenSerifBold\scriptscriptfont\bffam=\fiveSerifBold%
\def\cal{\tenSymbols}%
\def\greekbold{\tenMathBold}%
\def\gothic{\tenGothic}%
\def\Bbb{\tenDouble}%
\def\LieFont{\tenSerifItalics}%
\nt\normalbaselines\baselineskip=14pt%
}

\def\AddressStyle{\parindent=0pt\parskip=0pt\normalbaselineskip=14pt%
\def\nt{\eightSerif}%
\def\rm{\fam0\eightSerif}%
\textfont0=\eightSerif\scriptfont0=\sevenSerif\scriptscriptfont0=\fiveSerif
\textfont1=\eightMath\scriptfont1=\sevenMath\scriptscriptfont1=\fiveMath
\textfont2=\eightSymbols\scriptfont2=\sevenSymbols\scriptscriptfont2=\fiveSymbols
\textfont3=\eightMoreSymbols\scriptfont3=\sevenMoreSymbols\scriptscriptfont3=\fiveMoreSymbols
\textfont\itfam=\eightSerifItalics\def\it{\fam\itfam\eightSerifItalics}%
\textfont\slfam=\eightSerifSlanted\def\sl{\fam\slfam\eightSerifSlanted}%
\textfont\ttfam=\eightTypewriter\def\tt{\fam\ttfam\eightTypewriter}%
\textfont\bffam=\eightSerifBold%
\def\bf{\fam\bffam\eightSerifBold}\scriptfont\bffam=\sevenSerifBold\scriptscriptfont\bffam=\fiveSerifBold%
\def\cal{\eightSymbols}%
\def\greekbold{\eightMathBold}%
\def\gothic{\eightGothic}%
\def\Bbb{\eightDouble}%
\def\LieFont{\eightSerifItalics}%
\nt\normalbaselines\baselineskip=14pt%
}

\def\AbstractStyle{\parindent=0pt\parskip=0pt\normalbaselineskip=12pt%
\def\nt{\eightSerif}%
\def\rm{\fam0\eightSerif}%
\textfont0=\eightSerif\scriptfont0=\sevenSerif\scriptscriptfont0=\fiveSerif
\textfont1=\eightMath\scriptfont1=\sevenMath\scriptscriptfont1=\fiveMath
\textfont2=\eightSymbols\scriptfont2=\sevenSymbols\scriptscriptfont2=\fiveSymbols
\textfont3=\eightMoreSymbols\scriptfont3=\sevenMoreSymbols\scriptscriptfont3=\fiveMoreSymbols
\textfont\itfam=\eightSerifItalics\def\it{\fam\itfam\eightSerifItalics}%
\textfont\slfam=\eightSerifSlanted\def\sl{\fam\slfam\eightSerifSlanted}%
\textfont\ttfam=\eightTypewriter\def\tt{\fam\ttfam\eightTypewriter}%
\textfont\bffam=\eightSerifBold%
\def\bf{\fam\bffam\eightSerifBold}\scriptfont\bffam=\sevenSerifBold\scriptscriptfont\bffam=\fiveSerifBold%
\def\cal{\eightSymbols}%
\def\greekbold{\eightMathBold}%
\def\gothic{\eightGothic}%
\def\Bbb{\eightDouble}%
\def\LieFont{\eightSerifItalics}%
\nt\normalbaselines\baselineskip=12pt%
}

\def\RefsStyle{\parindent=0pt\parskip=0pt%
\def\nt{\eightSerif}%
\def\rm{\fam0\eightSerif}%
\textfont0=\eightSerif\scriptfont0=\sevenSerif\scriptscriptfont0=\fiveSerif
\textfont1=\eightMath\scriptfont1=\sevenMath\scriptscriptfont1=\fiveMath
\textfont2=\eightSymbols\scriptfont2=\sevenSymbols\scriptscriptfont2=\fiveSymbols
\textfont3=\eightMoreSymbols\scriptfont3=\sevenMoreSymbols\scriptscriptfont3=\fiveMoreSymbols
\textfont\itfam=\eightSerifItalics\def\it{\fam\itfam\eightSerifItalics}%
\textfont\slfam=\eightSerifSlanted\def\sl{\fam\slfam\eightSerifSlanted}%
\textfont\ttfam=\eightTypewriter\def\tt{\fam\ttfam\eightTypewriter}%
\textfont\bffam=\eightSerifBold%
\def\bf{\fam\bffam\eightSerifBold}\scriptfont\bffam=\sevenSerifBold\scriptscriptfont\bffam=\fiveSerifBold%
\def\cal{\eightSymbols}%
\def\greekbold{\eightMathBold}%
\def\gothic{\eightGothic}%
\def\Bbb{\eightDouble}%
\def\LieFont{\eightSerifItalics}%
\nt\normalbaselines\baselineskip=10pt%
}



%
%


\def\ModeYes{yes}
\def\ModeNo{no}

\def\ModeUndef{undefined}


\def\nx{\noexpand}
\def\ni{\noindent}
\def\newpage{\vfill\eject}

\def\ss{\vskip 5pt}
\def\ms{\vskip 10pt}
\def\bs{\vskip 20pt}

 \def\,{\mskip\thinmuskip}
 \def\!{\mskip-\thinmuskip}
 \def\>{\mskip\medmuskip}
 \def\;{\mskip\thickmuskip}

%
% ... 
%

\def\refsModePost{post}
\def\refsModeAuto{auto}

\def\dbRefsSatusModeOk{ok}
\def\dbRefsSatusModeError{error}
\def\dbRefsSatusModeWarning{warning}


\newcount\BNUM
\BNUM=0

\def\refs{}

\def\SetModePost{\xdef\refsMode{\refsModePost}}			
\SetModePost

\def\dbRefsStatusOk{%
	\xdef\dbRefsStatus{\dbRefsSatusModeOk}%
	\xdef\dbRefsError{\ModeNo}%
	\xdef\dbRefsWarning{\ModeNo}%
	\xdef\dbRefsInfo{\ModeNo}%
}

\def\dbRefs{%
}

\def\dbRefsGet#1{%
	\xdef\found{N}\xdef\ikey{#1}\dbRefsStatusOk%
	\xdef\key{\ModeUndef}\xdef\tag{\ModeUndef}\xdef\tail{\ModeUndef}%
	\dbRefs%
}

\def\NextRefsTag{%
	\global\advance\BNUM by 1%
}
\def\ShowTag#1{{\bf [#1]}}

\def\dbRefsInsert#1#2{%
\dbRefsGet{#1}%
\if\found Y %
   \xdef\dbRefsStatus{\dbRefsSatusModeWarning}%
   \xdef\dbRefsWarning{record is already there}%
   \xdef\dbRefsInfo{record not inserted}%
\else%
   \toks2=\expandafter{\dbRefs}%
   \ifx\refsMode\refsModeAuto \NextRefsTag
    \xdef\dbRefs{%
   	\the\toks2 \nx\xdef\nx\dbx{#1}%
	\nx\ifx\nx\ikey %
		\nx\dbx\nx\xdef\nx\found{Y}%
		\nx\xdef\nx\key{#1}%
		\nx\xdef\nx\tag{\the\BNUM}%
		\nx\xdef\nx\tail{#2}%
	\nx\fi}%
	\global\xdef\refs{\refs \ss\ni[\the\BNUM]\ #2\par}
   \fi%
   \ifx\refsMode\refsModePost 
    \xdef\dbRefs{%
   	\the\toks2 \nx\xdef\nx\dbx{#1}%
	\nx\ifx\nx\ikey %
		\nx\dbx\nx\xdef\nx\found{Y}%
		\nx\xdef\nx\key{#1}%
		\nx\xdef\nx\tag{\ModeUndef}%
		\nx\xdef\nx\tail{#2}%
	\nx\fi}%
   \fi%
\fi%
}

\def\dbRefsEdit#1#2#3{\dbRefsGet{#1}%
\if\found N 
   \xdef\dbRefsStatus{\dbRefsSatusModeError}%
   \xdef\dbRefsError{record is not there}%
   \xdef\dbRefsInfo{record not edited}%
\else%
   \toks2=\expandafter{\dbRefs}%
   \xdef\dbRefs{\the\toks2%
   \nx\xdef\nx\dbx{#1}%
   \nx\ifx\nx\ikey\nx\dbx %
	\nx\xdef\nx\found{Y}%
	\nx\xdef\nx\key{#1}%
	\nx\xdef\nx\tag{#2}%
	\nx\xdef\nx\tail{#3}%
   \nx\fi}%
\fi%
}

\def\bib#1#2{\RefsStyle\dbRefsInsert{#1}{#2}%
	\ifx\dbRefsStatus\dbRefsSatusModeWarning %
		\message{^^J}%
		\message{WARNING: Reference [#1] is doubled.^^J}%
	\fi%
}

\def\ref#1{\dbRefsGet{#1}%
\ifx\found N %
  \message{^^J}%
  \message{ERROR: Reference [#1] unknown.^^J}%
  \ShowTag{??}%
\else%
	\ifx\tag\ModeUndef \NextRefsTag%
		\dbRefsEdit{#1}{\the\BNUM}{\tail}%
		\dbRefsGet{#1}%
		\global\xdef\refs{\refs \ss\ni [\tag]\ \tail\par}
	\fi
	\ShowTag{\tag}%
\fi%
}

\def\ShowBiblio{\ms\Ensure{\SectionEnsure}%
{\SectionStyle\ni References}%
{\RefsStyle\refs}%
}

\newcount\CHANGES
\CHANGES=0
\def\AuxFile{7}
\def\PreventDoubleOn{\xdef\PreventDoubleLabel{\ModeYes}}

\PreventDoubleOn

\def\StoreLabel#1#2{\xdef\itag{#2}
 \ifx\PreModeStatus\ModeNo %
   \message{^^J}%
   \errmessage{You can't use Check without starting with OpenPreMode (and finishing with ClosePreMode)^^J}%
 \else%
   \immediate\write\AuxFile{\nx\dbLabelPreInsert{#1}{\itag}}%
   \dbLabelGet{#1}%
   \ifx\itag\tag %
   \else%
	\global\advance\CHANGES by 1%
 	\xdef\itag{(?.??)}%
    \fi%
   \fi%
}

\def\PreModeStatus{\ModeNo}

\def\ClosePreMode{\immediate\closeout\AuxFile%
  \ifnum\CHANGES=0%
	\message{^^J}%
	\message{**********************************^^J}%
	\message{**  NO CHANGES TO THE AuxFile  **^^J}%
	\message{**********************************^^J}%
 \else%
	\message{^^J}%
	\message{**************************************************^^J}%
	\message{**  PLAEASE TYPESET IT AGAIN (\the\CHANGES)  **^^J}%
    \errmessage{**************************************************^^ J}%
  \fi%
  \edef\PreModeStatus{\ModeNo}%
}

\def\dbLabelSatusModeOk{ok}

\def\dbLabelSatusModeWarning{warning}

\def\dbLabelStatusOk{%
	\xdef\dbLabelStatus{\dbLabelSatusModeOk}%
	\xdef\dbLabelError{\ModeNo}%
	\xdef\dbLabelWarning{\ModeNo}%
	\xdef\dbLabelInfo{\ModeNo}%
}

\def\dbLabel{%
}

\def\dbLabelGet#1{%
	\xdef\found{N}\xdef\ikey{#1}\dbLabelStatusOk%
	\xdef\key{\ModeUndef}\xdef\tag{\ModeUndef}\xdef\pre{\ModeUndef}%
	\dbLabel%
}

\def\ShowLabel#1{%
 \dbLabelGet{#1}%
 \ifx\tag \ModeUndef %
 	\global\advance\CHANGES by 1%
 	(?.??)%
 \else%
 	\tag%
 \fi%
}

\def\dbLabelPreInsert#1#2{\dbLabelGet{#1}%
\if\found Y %
  \xdef\dbLabelStatus{\dbLabelSatusModeWarning}%
   \xdef\dbLabelWarning{Label is already there}%
   \xdef\dbLabelInfo{Label not inserted}%
   \message{^^J}%
   \errmessage{Double pre definition of label [#1]^^J}%
\else%
   \toks2=\expandafter{\dbLabel}%
    \xdef\dbLabel{%
   	\the\toks2 \nx\xdef\nx\dbx{#1}%
	\nx\ifx\nx\ikey %
		\nx\dbx\nx\xdef\nx\found{Y}%
		\nx\xdef\nx\key{#1}%
		\nx\xdef\nx\tag{#2}%
		\nx\xdef\nx\pre{\ModeYes}%
	\nx\fi}%
\fi%
}

\def\dbLabelInsert#1#2{\dbLabelGet{#1}%
\xdef\itag{#2}%
\dbLabelGet{#1}%
\if\found Y %
	\ifx\tag\itag %
	\else%
	   \ifx\PreventDoubleLabel\ModeYes %
		\message{^^J}%
		\errmessage{Double definition of label [#1]^^J}%
	   \else%
		\message{^^J}%
		\message{Double definition of label [#1]^^J}%
	   \fi%
	\fi%
   \xdef\dbLabelStatus{\dbLabelSatusModeWarning}%
   \xdef\dbLabelWarning{Label is already there}%
   \xdef\dbLabelInfo{Label not inserted}%
\else%
   \toks2=\expandafter{\dbLabel}%
    \xdef\dbLabel{%
   	\the\toks2 \nx\xdef\nx\dbx{#1}%
	\nx\ifx\nx\ikey %
		\nx\dbx\nx\xdef\nx\found{Y}%
		\nx\xdef\nx\key{#1}%
		\nx\xdef\nx\tag{#2}%
		\nx\xdef\nx\pre{\ModeNo}%
	\nx\fi}%
\fi%
}


\newcount\PART
\newcount\CHAPTER
\newcount\SECTION
\newcount\SUBSECTION
\newcount\FNUMBER

\PART=0
\CHAPTER=0
\SECTION=0
\SUBSECTION=0	
\FNUMBER=0

\def\LastPart{\ModeUndef}
\def\LastChapter{\ModeUndef}
\def\LastSection{\ModeUndef}
\def\LastSubSection{\ModeUndef}
\def\LastClaim{\ModeUndef}
\def\Last{\ModeUndef}

\newdimen\TOBOTTOM
\newdimen\LIMIT

\def\Ensure#1{\ \par\ \immediate\LIMIT=#1\immediate\TOBOTTOM=\the\pagegoal\advance\TOBOTTOM by -\pagetotal%
\ifdim\TOBOTTOM<\LIMIT\newpage \else%
\vskip-\parskip\vskip-\parskip\vskip-\baselineskip\fi}

\def\PartLabel{\the\PART}
\def\NewPart#1{\global\advance\PART by 1%
         \bs\ni{\PartStyle  Part \PartLabel:}
         \bs\ni{\PartStyle #1}\newpage%
         \CHAPTER=0\SECTION=0\SUBSECTION=0\FNUMBER=0%
         \gdef\Left{#1}%
         \global\edef\Last{\PartLabel}%
         \global\edef\LastPart{\PartLabel}%
         \global\edef\LastChapter{\ModeUndef}%
         \global\edef\LastSection{\ModeUndef}%
         \global\edef\LastSubSection{\ModeUndef}%
         \global\edef\LastClaim{\ModeUndef}}
\def\ChapterLabel{\the\CHAPTER}
\def\NewChapter#1{\global\advance\CHAPTER by 1%
         \bs\ni{\ChapterStyle  Chapter \ChapterLabel: #1}\ms%
         \SECTION=0\SUBSECTION=0\FNUMBER=0%
         \gdef\Left{#1}%
         \global\edef\Last{\ChapterLabel}%
         \global\edef\LastChapter{\ChapterLabel}%
         \global\edef\LastSection{\ModeUndef}%
         \global\edef\LastSubSection{\ModeUndef}%
         \global\edef\LastClaim{\ModeUndef}}
\def\SectionEnsure{3cm}
\def\NewSection#1{\Ensure{\SectionEnsure}\gdef\SectionLabel{\the\SECTION}\global\advance\SECTION by 1%
         \ms\ni{\SectionStyle  \SectionLabel.\ #1}\ss%
         \SUBSECTION=0\FNUMBER=0%
         \gdef\Left{#1}%
         \global\edef\Last{\SectionLabel}%
         \global\edef\LastSection{\SectionLabel}%
         \global\edef\LastSubSection{\ModeUndef}%
         \global\edef\LastClaim{\ModeUndef}}
\def\NewAppendix#1#2{\Ensure{\SectionEnsure}\gdef\SectionLabel{#1}\global\advance\SECTION by 1%
         \bs\ni{\SectionStyle  Appendix \SectionLabel.\ #2}\ss%
         \SUBSECTION=0\FNUMBER=0%
         \gdef\Left{#2}%
         \global\edef\Last{\SectionLabel}%
         \global\edef\LastSection{\SectionLabel}%
         \global\edef\LastSubSection{\ModeUndef}%
         \global\edef\LastClaim{\ModeUndef}}
\def\Acknowledgements{\Ensure{\SectionEnsure}\gdef\SectionLabel{}%
         \ms\ni{\SectionStyle  Acknowledgments}\ss%
         \SECTION=0\SUBSECTION=0\FNUMBER=0%
         \gdef\Left{}%
         \global\edef\Last{\ModeUndef}%
         \global\edef\LastSection{\ModeUndef}%
         \global\edef\LastSubSection{\ModeUndef}%
         \global\edef\LastClaim{\ModeUndef}}
\def\SubSectionEnsure{2cm}
\def\SubSectionLabel{\ifnum\SECTION>0 \the\SECTION.\fi\the\SUBSECTION}
\def\NewSubSection#1{\Ensure{\SubSectionEnsure}\global\advance\SUBSECTION by 1%
         \ms\ni{\SubSectionStyle #1}\ss%
         \global\edef\Last{\SubSectionLabel}%
         \global\edef\LastSubSection{\SubSectionLabel}}
\def\SetNumberingModeN{\def\ClaimLabel{(\the\FNUMBER)}}
\def\SetNumberingModeSN{\def\ClaimLabel{(\ifnum\SECTION>0 \SectionLabel.\fi%
      \the\FNUMBER)}}
\def\SetNumberingModeCSN{\def\ClaimLabel{(\ifnum\CHAPTER>0 \the\CHAPTER.\fi%
      \ifnum\SECTION>0 \SectionLabel.\fi%
      \the\FNUMBER)}}

\def\NewClaim{\global\advance\FNUMBER by 1%
    \ClaimLabel%
    \global\edef\LastClaim{\ClaimLabel}%
    \global\edef\Last{\ClaimLabel}}

\def\HideLabels{\xdef\ShowLabelsMode{\ModeNo}}
\HideLabels

\def\fn{\eqno{\NewClaim}} 
\def\fl#1{%
\ifx\ShowLabelsMode\ModeYes%
 \eqno{{\buildrel{\hbox{\AbstractStyle[#1]}}\over{\hfill\NewClaim}}}%
\else%
 \eqno{\NewClaim}%
\fi%
\dbLabelInsert{#1}{\ClaimLabel}}
\def\fprel#1{\global\advance\FNUMBER by 1\StoreLabel{#1}{\ClaimLabel}%
\ifx\ShowLabelsMode\ModeYes%
\eqno{{\buildrel{\hbox{\AbstractStyle[#1]}}\over{\hfill.\itag}}}%
\else%
 \eqno{\itag}%
\fi%
}

\def\cl#1{\global\advance\FNUMBER by 1\dbLabelInsert{#1}{\ClaimLabel}%
\ifx\ShowLabelsMode\ModeYes%
${\buildrel{\hbox{\AbstractStyle[#1]}}\over{\hfill\ClaimLabel}}$%
\else%
  $\ClaimLabel$%
\fi%
}
\def\cprel#1{\global\advance\FNUMBER by 1\StoreLabel{#1}{\ClaimLabel}%
\ifx\ShowLabelsMode\ModeYes%
${\buildrel{\hbox{\AbstractStyle[#1]}}\over{\hfill.\itag}}$%
\else%
  $\itag$%
\fi%
}

\def\Note{\ms\leftskip 3cm\rightskip 1.5cm\AbstractStyle}
\def\endNote{\par\leftskip 2cm\rightskip 0cm\NormalStyle\ss}


\parindent=7pt
\leftskip=2cm
\newcount\SideIndent
\newcount\SideIndentTemp
\SideIndent=0
\newdimen\SectionIndent
\SectionIndent=-8pt

\def\sidebar{\vrule height15pt width.2pt }
\def\endcorner{\hbox{\hbox{\vrule height6pt width.2pt}\vbox to6pt{\vfill\hbox
to4pt{\leaders\hrule height0.2pt\hfill}}}}
\def\begincorner{\hbox{\hbox{\vrule height6pt width.2pt}\vbox to6pt{\hbox
to4pt{\leaders\hrule height0.2pt\hfill}}}}
\def\endbegincorner{\hbox{\vbox to15pt{\endcorner\vskip-6pt\begincorner\vfill}}}
\def\SideShow{\SideIndentTemp=\SideIndent \ifnum \SideIndentTemp>0 
\loop\sidebar\hskip 2pt \advance\SideIndentTemp by-1\ifnum \SideIndentTemp>1 \repeat\fi}

\def\BeginSection{{\vbadness 100000 \par\ni\hskip\SectionIndent%
\SideShow\vbox to 15pt{\vfill\begincorner}}\global\advance\SideIndent by1\vskip-10pt}

\def\EndSection{{\vbadness 100000 \par\ni\global\advance\SideIndent by-1%
\hskip\SectionIndent\SideShow\vbox to15pt{\endcorner\vfill}\vskip-10pt}}

\def\EndBeginSection{{\vbadness 100000\par\ni%
\global\advance\SideIndent by-1\hskip\SectionIndent\SideShow
\vbox to15pt{\vfill\endbegincorner}}%
\global\advance\SideIndent by1\vskip-10pt}

\def\ShowBeginCorners#1{%
\SideIndentTemp =#1 \advance\SideIndentTemp by-1%
\ifnum \SideIndentTemp>0 %
\vskip-15truept\hbox{\kern 2truept\vbox{\hbox{\begincorner}%
\ShowBeginCorners{\SideIndentTemp}\vskip-3truept}}%
\fi%
}

\def\ShowEndCorners#1{%
\SideIndentTemp =#1 \advance\SideIndentTemp by-1%
\ifnum \SideIndentTemp>0 %
\vskip-15truept\hbox{\kern 2truept\vbox{\hbox{\endcorner}%
\ShowEndCorners{\SideIndentTemp}\vskip 2truept}}%
\fi%
}

\def\BeginSections#1{{\vbadness 100000 \par\ni\hskip\SectionIndent%
\SideShow\vbox to 15pt{\vfill\ShowBeginCorners{#1}}}\global\advance\SideIndent by#1\vskip-10pt}

\def\EndSections#1{{\vbadness 100000 \par\ni\global\advance\SideIndent by-#1%
\hskip\SectionIndent\SideShow\vbox to15pt{\vskip15pt\ShowEndCorners{#1}\vfill}\vskip-10pt}}

\def\EndBeginSections#1#2{{\vbadness 100000\par\ni%
\global\advance\SideIndent by-#1%
\hbox{\hskip\SectionIndent\SideShow\kern-2pt%
\vbox to15pt{\vskip15pt\ShowEndCorners{#1}\vskip4pt\ShowBeginCorners{#2}}}}%
\global\advance\SideIndent by#2\vskip-10pt}




%
%


\def\al{\alpha}
\def\be{\beta}
\def\de{\delta}

\def\ep{\epsilon}

\def\la{\lambda}

\def\om{\omega}
\def\si{\sigma}
\def\vp{\varphi}

\def\Ga{\Gamma}

\def\La{\Lambda}


 \def\calB{{\hbox{\cal B}}}
 \def\calC{{\hbox{\cal C}}}
 
 \def\calP{{\hbox{\cal P}}}
 \def\calL{{\hbox{\cal L}}}

 \def\calR{{\hbox{\cal R}}}


 \def\gotC{{\hbox{\gothic C}}}
 \def\gotP{{\hbox{\gothic P}}}
 


 \def\R{{\hbox{\Bbb R}}}

 \def\R{{\hbox{\Bbb R}}}


\def\Con{{\hbox{Con}}}

\def\Div{{\hbox{Div}}}

\def\Lor{{\hbox{Lor}}}

\def\id{{\hbox{\rm id}}}

\def\ip{\hbox to4pt{\leaders\hrule height0.3pt\hfill}\vbox to8pt{\leaders\vrule width0.3pt\vfill}\kern 2pt}
 
\def\del{\partial}
\def\na{\nabla}

\def\arr{\rightarrow}

%
%

\def\cases#1{\left\{\eqalign{#1}\right.}
\NormalStyle
\SetNumberingModeSN
\PreventDoubleOn

\long\def\title#1{\centerline{\TitleStyle\ni#1}}

\long\def\author#1{\ms\centerline{\AuthorStyle by {\it #1}}}

\long\def\address#1{\ss\centerline{\AddressStyle #1}\par}
\long\def\moreaddress#1{\centerline{\AddressStyle #1}\par}
\def\abstract{\ms\leftskip 3cm\rightskip .5cm\AbstractStyle{\bf \ni Abstract:}\ }
\def\endabstract{\par\leftskip 2cm\rightskip 0cm\NormalStyle\ss}

\SetNumberingModeSN

\def\frac[#1/#2]{\hbox{$#1\over#2$}}
\def\Frac[#1/#2]{{#1\over#2}}
\def\({\left(}
\def\){\right)}
\def\[{\left[}
\def\]{\right]}
\def\^#1{{}^{#1}_{\>\cdot}}
\def\_#1{{}_{#1}^{\>\cdot}}
\def\Label=#1{{\buildrel {\hbox{\fiveSerif \ShowLabel{#1}}}\over =}}
\def\<{\kern -1pt}

\def\Dal{\hbox{\tenRelazioni  \char003}}


\def\ExpandAllCNotes{\long\def\CNote##1{%
\BeginSection
	\Note%
 		##1%
	\endNote%
\EndSection%
}}
\ExpandAllCNotes
%
%
%
%


\def\frame#1{\vbox{\hrule\hbox{\vrule\vbox{\kern2pt\hbox{\kern2pt#1\kern2pt}\kern2pt}\vrule}\hrule\kern-4pt}}

\def\Box to #1#2#3{\frame{\vtop{\hbox to #1{\hfill #2 \hfill}\hbox to #1{\hfill #3 \hfill}}}}


\bib{EPS} {J. Elhers, F.A. E. Pirani, A. Schild, 
{\it The Geometry of free fall and light propagation in Studies in Relativity}, 
Papers in honour of J. L. Synge 6384 (1972)}

\bib{Schouten} {J.A.Schouten, 
{\it Ricci-Calculus: An Introduction to Tensor Analysis and its Geometrical Applications}, 
Springer Verlag (1954)}

\bib{Pons}{N. Dadhich, J.M. Pons, 
{\it Equivalence of the Einstein-Hilbert and the Einstein-Palatini formulations of general relativity for an arbitrary connection}, 
(to appear on GRG); arXiv:1010.0869v3 [gr-qc]}

\bib{TM1} {L. Fatibene, M.Ferraris, M. Francaviglia, G. Magnano,
{\it Extended Theories of Gravitation: Structure of Spacetime and Fundamental Principles of Physics, following Ehlers-Pirani-Schild Framework},
EPJ Web of Conferences 58, 02002 (2013); http://dx.doi.org/10.1051/epjconf/20135802002}

\bib{TM2}{L. Fatibene, M.Ferraris, M. Francaviglia, G. Magnano,
{\it Extended Theories of Gravitation: Observation Protocols and Experimental Tests},
EPJ Web of Conferences 58, 02007 (2013)
http://dx.doi.org/10.1051/epjconf/20135802007}

\bib{Geod}{L. Fatibene, M. Francaviglia, G. Magnano,
{\it  On a Characterization of Geodesic Trajectories and Gravitational Motions},
Int. J. Geom. Meth. Mod. Phys.; arXiv:1106.2221v2 [gr-qc]}

 \bib{Fabbri}{L.Fabbri,
{\it Conformal Gravity with Electrodynamics for Fermion Fields and their Symmetry Breaking Mechanism},
Int. J. Geom. Meth. Mod.Phys. 11, 1450019 (2014); arXiv:1205.5386 [gr-qc]
}

 \bib{FatiMatter}{L.Fatibene, M.Francaviglia, S. Mercadante,
{\it Matter Lagrangians Coupled with Connections},
Int. J. Geom. Methods Mod. Phys. {\bf 7}(5) (2010), 1185-1189; arXiv:0911.2981}

\bib{Univ}{A. Borowiec, M. Ferraris, M. Francaviglia, I. Volovich, 
{\it Universality of Einstein Equations for the Ricci Squared Lagrangians}, 
Class. Quantum Grav. 15, 43-55, 1998}

\bib{Magnano}{G. Magnano, L.M. Sokolowski, 
{\it On Physical Equivalence between Nonlinear Gravity Theories}
Phys.Rev. D50 (1994) 5039-5059; gr-qc/9312008}

\bib{Kepler}{L. Fatibene, M.Francaviglia,
{\it Mathematical Equivalence versus Physical Equivalence between Extended Theories of Gravitation.}
Int. J. Geom. Meth. Mod. Phys., {\bf 11}(01) (2013), 1450008; 	arXiv:1302.2938 [gr-qc]}

\bib{F1}{L. Fatibene, M. Francaviglia,
{\it Weyl Geometries and Timelike Geodesics},
Int. J. Geom. Methods Mod. Phys. 9(5) (2012) 1220006; arXiv:1106.1961v1 [gr-qc]
}

\bib{F3}{L. Fatibene, M.Francaviglia,
{\it Fluids in Weyl Geometries},
Int. J. Geom. Meth. Mod. Phys., {\bf 9}(2), (2009), 1260003}

\bib{Perlick}{V.Perlick,
{\it Characterization of Standard Clocks by Means of Light Rays and Freely Falling Particles},
Gen. Rel. Grav. {\bf 19}(11), (1987) 1059-1073
}

\bib{Polistina}{L. Fatibene, M. Polistina,
{\it Breaking the Conformal Gauge by Fixing Time Protocols},
(in preparation)
}

\bib{Mana}{A.Mana, L.Fatibene, M.Ferraris,
{\it A further study on Palatini $f (\calR)$-theories for polytropic stars},
(in preparation)
}

\bib{Olmo}{L.Fatibene and M.Francaviglia,
{\it Extended Theories of Gravitation and the Curvature of the Universe -- Do We Really Need Dark Matter?}
in : Open Questions in Cosmology, Edited by Gonzalo J. Olmo, Intech (2012), ISBN 978-953-51-0880-1; DOI: 10.5772/52041}

\bib{Sout}{E.Barausse, Thomas P.Sotiriou, J.C.Miller, 
{\it A no-go theorem for polytropic spheres in Palatini $f(R)$ gravity},
DOI: 10.1088/0264-9381/25/6/062001\ (4th March 2008)}

\bib{Olmo-No-go}{G.J.Olmo, 
{\it Re-examination of polytropic spheres in Palatini $f(R)$ gravity},
DOI: 10.1103/PhysRevD.78.104026\ (20th October 2008)}


\def\ubal{\underline{\al}\kern1pt}
\def\obal{\overline{\al}\kern1pt}

\def\ubR{\underline{R}\kern1pt}
\def\obR{\overline{R}\kern1pt}
\def\ubom{\underline{\om}\kern1pt}
\def\obxi{\overline{\xi}\kern1pt}
\def\ubu{\underline{u}\kern1pt}
\def\ube{\underline{e}\kern1pt}
\def\obe{\overline{e}\kern1pt}

\NormalStyle

\title{Extended Gravity\footnote{$^\dagger$}{\AbstractStyle This paper is dedicated to the memory of Mauro Francaviglia.}}

\author{L.Fatibene$^{a,b}$, S.Garruto$^{a,b}$}

\address{$^a$ Department of Mathematics, University of Torino (Italy)}
\moreaddress{$^b$ INFN - Sez.~Torino, Iniz.~Spec.~QGSKY}

\abstract
We shall show equivalence between Palatini-$f(\calR)$ theories and Brans-Dicke (BD)  theories at the level of action principles in generic dimension with generic matter coupling.
We do that by introducing the Helmholtz Lagrangian associated to Palatini-$f(\calR)$ theory and then performing frame transformations in order to recover Einstein frame and Brans-Dicke frame.
This clarifies the relation among different formulations and the transformations among different frames.
Additionally, it defines a formulation {\it a l\'a Palatini} for the Brans-Dicke theory which is dynamically equivalent to metric BD
(unlike the standard Palatini-formulation of metric BD theory which are {\it not} dynamically equivalent).

In conclusion we discuss interpretation of extended theories of gravitation and perspectives.
\endabstract

\NewSection{Introduction}

Standard General Relativity (GR) is a remarkably effective theory which modeled for almost a century the whole phenomenology of gravity we were able to observe.
During the last decade of $20^{th}$ century new and more precise  observations in astrophysics and cosmology started to be available.
Today we have a whole class of phenomena which can be fitted by standard GR at the price of introducing gravitational sources which do not correspond to any kind of matter we see around  us.  Such sources are collectively called {\it dark sources}.

Currently, it is estimated that the visible matter of the universe provides about $4\%$ of its total mass and energy. 
About $23\%$ is the so-called {\it dark matter} (in the current understanding made of particles which interact only through weak and gravitational interactions)
and about $73\%$ is made of so-called {\it dark energy} (in the current understanding best candidate is a tiny positive cosmological constant which should be identified with vacuum energy in standard model, though the order of magnitude does not quite match).

Although there are strong evidences of dark sources both from astrophysics and cosmology all evidences for them are purely gravitational effects. We could say that we know dark sources uniquely through their gravitational effect and their fundamental nature from particle physics is  unknown.

In particular, there is no candidate in the standard model of particle physics for dark matter  and one has to resort to still unobserved new particles of which there is no evidence at fundamental level, yet. Moreover, experimental survey which are trying to directly detect effects of dark matter at fundamental level (e.g.~gamma rays produced by annihilation or direct detection of weak interaction with ordinary matter) are (until now) simply not detecting anything.
Of course, we are in the beginning of researches in this field and situation could change at any moment. However, until now we have to remark our complete  ignorance about the nature of dark matter. Also observations are cutting down more and more candidates based on speculations about extensions of the standard model.

About dark energy the situation is similar. A good candidate is available as a tiny positive cosmological constant. However, in the current understanding cosmological constant should be related to the quantum vacuum expectation value of energy which strongly depends on the details of the model.
The quantum vacuum expectation value of standard model seems too big to account for gravitational effects of dark energy, string theory models usually give the wrong sign for it. Of course, it is possible that still un-investigated models with the correct prediction exist. However, one has to again remark our ignorance about details. 
Moreover, it is still unclear whether observations point to a cosmological constant or to a so-called {\it quintessence field} (which is quite similar to a scalar field which plays the role of a cosmological constant though being allowed to depend on points in spacetime).
This is the quite discouraging state of our knowledge.

An alternative approach is to try and regard dark sources as purely gravitational effects due to slight modification of long range gravity interaction and dynamics. Also in this direction some success is available. For example, there are known dynamics for fitting different situations (e.g.~rotation profiles of galaxies, structure formations, cosmological acceleration) though currently there is no accepted modification which is able to account for all dark effects at all scales. 

Here we shall discuss some approaches  in the second line of reasoning. Unfortunately, once one allows modifications of gravitational interactions the range of possible modifications often becomes soon very wide. Moreover, interpretation of standard GR and observations very often is extremely dependent on the details of the model and more general models often need a scrutiny of foundations and interpretation of the theory.
This of course is particularly interesting since it gives us a better founded interpretation of gravitational theories and astronomical observations, even if in the end observations will point towards standard GR.
Probably, one should also keep in mind that when standard GR was developed a clear and definite  knowledge of connections was still on the way and the standard interpretation is strongly based on the assumption that the theory is governed by essentially one metric field which plays the multiple roles of gravitational field, geometry of spacetime, geometric counterpart of observational protocols, governing free fall of test particles and light propagation.

\ms
We shall hereafter consider a particular subclass of modified models for gravity, known as Palatini-$f(\calR)$ theories. We shall show how the physical interpretation of the theory is particularly sensitive to the detail of the model. 
 
In Section 2 we shall briefly review Ehlers-Pirani-Schild (EPS) framework for interpretation of gravitational theories and define the class of {\it extended theories of gravitation (ETG)}.
In Section  3 we shall review notation and main results about Palatini-$f(\calR)$ theories.
In Section 4 we shall present and discuss the equivalence with Brans-Dicke theories and behavior with respect to conformal transformations from a variational viewpoint.
In conclusions we shall discuss a tentative model and discuss possible consequences.

\NewSection{EPS framework}

In the early 70s Ehlers, Pirani and Schild (EPS) proposed an axiomatic framework for relativistic theories in which they showed how
one can derive the geometric structure of spacetime from potentially observable quantities, i.e.~worldlines of particles and light rays; see \ref{EPS}.
Accordingly, in the EPS framework the geometry of spacetime is not assumed but derived from more fundamental objects.

By assuming two congruences of worldlines  for particles ($\calP$) and light rays ($\calL$) on a spacetimes manifold $M$,
one can define out of light rays  a {\it conformal class} of metrics $\gotC=[g]$.
Two metrics $\tilde g$, $g\in \Lor(M)$ are {\it conformally equivalent} 
iff there exists a positive scalar field $\vp$ such that  $\tilde g= \vp(x)\cdot g$.
Let us stress that conformal transformations are {\it vertical transformations} on the configuration bundle $\Lor(M)$.

Then one can prove that  particles free fall is described by a {\it projective class} $\gotP=[\tilde\Ga]$ of connections; see \ref{Geod}.
Two connections are projectively equivalent iff they share the same autoparallel trajectories. In fact the connection
$\tilde\Ga'^\al_{\be\mu}= \tilde\Ga^\al_{\be\mu} + \de^\al_{(\be} V_{\mu)}$
defines the same geodesic trajectories as $\tilde\Ga^\al_{\be\mu}$ for any covector $V_\mu$.
In this case we say that $\tilde\Ga$ and $\tilde\Ga'$ are {\it projectively equivalent}. 
Accordingly, free fall corresponds to a projective class $\gotP=[\tilde\Ga]$; see \ref{Schouten}.

Finally, we need a compatibility condition between the conformal class $\gotC$ associated to light cones and the projective class $\gotP$ associated to free fall.
This is due by the simple fact that we know that light rays (and hence light cones) feel the gravitational field as mass particles.
Noticing that $g$-lightlike $g$-geodesics are conformally invariant (unlike general $g$-geodesics), we have then to assume that  $g$-lightlike $g$-geodesics
are a proper subset of all $\tilde\Ga$-autoparallel trajectories.
In view of EPS-compatibility condition one can show that a representative $\tilde\Ga\in \gotP$ of the projective structure can  always (and uniquely) be chosen
so that there exists a covector $A=A_\mu\>dx^\mu$ such that
$\tilde\na g= 2A\otimes g$
where $g\in \gotC$ is a representative of the conformal structure  and the covariant derivative $\tilde\na$  is the one uniquely associated to $\tilde\Ga$; see \ref{Pons}.
Equivalently one has the following relation between a representative $g$  of the conformal structure and the  representative $\tilde \Ga$ of the projective structure: 
$$\tilde\Ga^\al_{\be\mu}= \{g\}^\al_{\be\mu} + (g^{\al\ep} g_{\be\mu} -2\de^\al_{(\be}\de^\ep_{\mu)})A_\ep
\fl{compatibilityCondition}$$
To summarize, by assuming compatibility between particles and light rays one can define on spacetime a {\it EPS structure}, i.e.~a triple $(M, \gotC, \gotP)$.
The conformal structure $\gotC$ describes light cones and it is associated to light rays. Notice that having just a conformal structure one cannot yet define distances (which are not conformally invariant). Not being gauge covariant, in order to define distances one must resort to a {\it convention} which corresponds to the choice of a specific representative $g\in \gotC$. 
On the other hand, the projective structure $\gotP$ is associated to free fall so that one can make a canonical gauge fixing by choosing the only representative in the form \ShowLabel{compatibilityCondition} or, equivalently, the $1$-form $A$.

The triple $(M, \gotC, \tilde\Ga)$  (or, equivalently, the triple $(M, \gotC, A)$) is called a {\it Weyl geometry} on spacetime.
This setting is more general than the setting for standard GR where one has just a Lorentzian metric $ g$  determining both the conformal structure 
$ g\in \gotC$ and the free fall $\tilde\Ga=\{ g\}$ (i.e.~assuming that the connection $\tilde\Ga$ is obtained as the Levi-Civita connection uniquely associated to $g$).
Hence  standard GR is a simple and very peculiar case of EPS framework, where there is a gauge fixing of the conformal gauge. 
Such a fixing is possible iff the covector $A=A_\mu dx^\mu$ is exact, i.e.~$A=d\al$.
In this case, there exists a Lorentzian metric $\tilde g\in \gotC$ also determining free fall by $\tilde \Ga=\{\tilde g\}$.
When this happens the Weyl geometry $(M, [\tilde g], \{\tilde g\})$ is called an {\it integrable Weyl geometry}.
Notice that this is still more general than standard GR in the sense that the metric determining free fall and light cones is not the original $g$ chosen to describe dynamics, but a conformally related one  $\tilde g\in [g]$.
Reverting to standard GR in a sense amounts to choose the gauge $\al$ to be a constant (so that $A$ vanishes identically).

In integrable Weyl geometries there is nothing ensuring that the canonical representative $\tilde g$ also gives us the measured distances,
that as far as we can see could as well be related to any other conformally equivalent metric $g$.
Fixing the metric that we use to calculate distances is, at the end, a choice that we can do only {\it a posteriori}, on the basis of observations,
as Riemann clained as early as 1854: {\it the curvature of the universe has to be determined by astronomical observations}.
Deciding which is the metric that really enters observational protocols is something that should not be imposed {\it a priori} but rather something to be tested locally.

In view of EPS framework it is natural to assume that gravitational field is described by a metric $g$, which is the one which encodes geometrically the observational protocols and represents the conformal structure $\gotC$, and a (torsionless) connection $\tilde\Ga$ which is related to free fall of test particles. 

In fact, the request to be torsionless can be removed. If torsion is allowed, since the connection is fixed by free fall of test particles and geodesics equation does not depend on the torsion, arbitrary torsion can be introduced and equation \ShowLabel{compatibilityCondition} just determines the torsionless part of the connection. Of course torsion can become important when coupling to spinors (see \ref{Fabbri}). Here we restrict to torsionless connections for simplicity.

Then dynamics will determine a relation between these two structures. If dynamics enforces EPS-compatibility then the theory is called an {\it extended theory of gravitation (ETG)} and EPS-framework is implemented in the field theory.
A class of examples of ETG is provided by Palatini-$f(\calR)$ theories (see next Section). 
We shall see that generically, provided that matter couples to $g$ and not to $\tilde\Ga$, these theories are always integrable ETG since their dynamics forces the connection to be a metric connection for a representative $\tilde g$ of the conformal class, usually different from the one $g$ assumed to describe observational protocols.

Studying Palatini-$f(\calR)$ theories is therefore interesting since it allows to review foundations and interpretation of gravitational theories in a more general context.
Of course, Palatini-$f(\calR)$ theories are not the most general ETG; see \ref{FatiMatter}.

\NewSection{Palatini-$f(\calR)$ theories}

In Palatini-$f(\calR)$ theories  fundamental fields are a Lorentzian metric $g_{\mu\nu}$ and a connection $\tilde\Ga^\al_{\be\mu}$ assumed to be torsionless.
Let then $M$ be a smooth orientable connected paracompact manifold of dimension $m$, which can support (global) Lorentzian metrics. 
The configuration bundle will thence be 
$$
\calC= \Lor(M)\times_M \Con(M)\times_M B
\fn$$
where $\Lor(M)$ is the bundle of Lorentzian metrics (which allows global sections by assumption), $\Con(M)$ is the bundle of (torsionless) connections over $M$, and $B$ is a configuration bundle for matter fields (which, for example,  may include electromagnetic field and other matter fields).
The configuration bundle has local fibered coordinates $(x^\mu, g_{\mu\nu}, \tilde \Ga^\al_{\be\mu}, \psi^i)$.

The connection defines a Ricci tensor $\tilde R_{\al\be}(\tilde \Ga) $ (which is not necessarily symmetric in its indices) and let us set $\calR:=g^{\al\be} \tilde R_{\al\be}$ for a scalar curvature which depends on both the connection and the metric. 
The phase bundle is assumed to be 
$$
J\calC= J^1\Lor(M)\times_M J^1\Con(M)\times_M J^1 B
\fn$$
since the matter fields can couple to $g$ through the Levi-Civita connection $\{g\}$.

The dynamics is obtained from a Lagrangian in the form
$$
L= \[\sqrt{g} f(\calR)  + \calL_m(g, \psi)\] d\si
\fl{Lag}$$
where $\sqrt{g}$ denotes the square root of the absolute value of the determinant of the metric, $f(\calR)$ is a generic analytic (or sufficiently regular)
function of the scalar curvature $\calR$. For now we shall not assume much about the matter sector except that it does not couple to $\tilde\Ga$.

The Lagrangian can be varied with respect to fields to obtain
$$
\de L= \sqrt{g}(f' \tilde R_{\al\be} -\frac[1/2] f g_{\al\be} - T_{\al\be}) \de g^{\al\be} -\tilde \na_\la (\sqrt{g} f' g^{\al\be}) \de \tilde u^\la_{\al\be} + E_i\de \psi^i
+(\Div) 
\fn$$
where $T_{\al\be}$ is the Hilbert stress tensor of matter fields obtained by variation of the matter Lagrangian with respect to the metric field, we set 
$\tilde u^\la_{\al\be}:= \tilde \Ga^\la_{\al\be} - \de^\la_{(\al} \tilde \Ga^\ep_{\be)\ep}$, $\tilde\na$ denotes the covariant derivative induced by $\tilde \Ga$ and $(\Div)$ denotes pure divergences which do not affect field equations. 
Field equations induced by Lagrangian \ShowLabel{Lag} are then in the form
$$
\cases{
&f' \tilde R_{(\al\be)} -\frac[1/2] f g_{\al\be} = T_{\al\be}\cr
&\tilde \na_\la (\sqrt{g} f' g^{\al\be})=0\cr
&E_i=0\cr
}
\fl{fREQ}$$
These are second order equations in the connection (due to the $f'(\calR)$ factor in the second equation).
However one can define a new conformal metric $\tilde g_{\mu\nu}= \vp g_{\mu\nu}$ by using the conformal factor $\vp=(f'(\calR))^{\frac[2/m-2]}$ so that the second equation can be solve explicitly to give 
%
$$
\tilde \Ga^\al_{\be\mu}=\{\tilde g\}^\al_{\be\mu}= \{ g\}^\al_{\be\mu}-\frac[1/2]\(g^{\al\ep}g_{\be\mu} - 2\de^\al_{(\be}\de^\ep_{\mu)}\)\na_\ep \ln \vp
\fn$$
This shows that dynamics enforces EPS-compatibility and  that the projective structure is integrable.
With this information we can set $\tilde R_{\al\be}$ to be the Ricci tensor of the conformal metric $\tilde g$.
One can trace the first equation by $g^{\al\be}$ to obtain the so-called {\it master equation}
$$
f'(\calR)  \calR -\frac[m/2] f(\calR) = T
\fn$$ 
which is an algebraic equation for $\calR$ with a parameter $T:= g^{\al\be}T_{\al\be}$.

If $f''(\calR)  \calR +\(1 -\frac[m/2] \) f'(\calR)\not= 0$ (i.e.~except in the cases $f(\calR)=C \calR^{\frac[m/2]} + c$)
then one can generically solve for $\calR=\calR(T)$.
Another important function is the one obtained by inverting  $\vp=(f'(\calR))^{\frac[2/m-2]}$ for $\calR$ to obtain $\calR=r(\vp)$.

In any event the first field equation can be recast in the form
$$
\tilde R_{\al\be}=\vp^{-\frac[m-2/2]} \(T_{\al\be}-\frac[1/m]Tg_{\al\be}\) + \frac[1/m ]  \calR(T) g_{\al\be}
\fl{FEQ1}$$

From this one can see that Palatini-$f(\calR)$ theories are equivalent to standard GR for the conformal metric $\tilde g$ though with a modified source stress tensor.
In fact one has
$$
\tilde G_{\al\be}:=\tilde R_{\al\be} - \frac[1/2]\tilde R \tilde g_{\al\be}= \vp^{-\frac[m-2/2]} \(T_{\al\be}-\frac[1/m]Tg_{\al\be}\) - \frac[m-2/2m ]  \calR(T) g_{\al\be}=: \tilde T_{\al\be}
\fn$$

In vacuum, $T_{\al\be}=0$ and $\Lambda:= - \frac[m-2/2m ]  \calR(0)$ is constant, so that Palatini-$f(\calR)$ theories is equivalent to standard GR with a cosmological constant $\La$; see \ref{Univ}.

\NewSubSection{Brans-Dicke theories with potential}

Let us consider a metric $g$, a scalar field $\vp$ and generic matter fields $\psi^i$.
A Brans-Dicke theory  (in dimension $m$) is described by a Lagrangian of the form
$$
L_{BD}= \[\sqrt{g} \(\vp^\al R -\frac[\om/\vp^\be] \na_\mu \vp \na^\mu\vp+ U(\vp)\) + \calL_m (g, \psi)\] d\si
\fn$$
where $R$ is of course the scalar curvature of $g$, $\al, \be, \om$ are constants and $U(\vp)$ a potential describing self interactions of the scalar field $\vp$.

One can easily check that field equations for the Brans-Dicke Lagrangian are
$$
\cases{
&\vp^\al G_{\mu\nu} = T_{\mu\nu}  + \al\vp^{\al-1}\(\na_{\mu\nu}\vp -\Dal\vp g_{\mu\nu} \) 
+\(\al(\al-1)\vp^{\al-2} + \frac[\om/\vp^\be]\) \na_\mu\vp\na_\nu \vp +\cr
&\qquad\qquad
	+\frac[\al(\al-1)/m-2]\vp^{\al-2}\na_\la\vp\na^\la\vp g_{\mu\nu}
-\frac[m-2/2] U g_{\al\be}\cr
&\al\vp^\al R- \frac[\be\om/\vp^\be]\na_\la\vp\na^\la\vp + \frac[2\om/\vp^{\be-1}]\Dal\vp +\vp U'=0	\cr
&E_i=0\cr
}
\fn$$
One can subtract from the first equation its trace and eliminate the scalar curvature by comparing the trace of the first and the second equation, so that field equations can be recast as
$$
\cases{
&\vp^\al R_{\mu\nu} = T_{\mu\nu}  -\frac[1/2]Tg_{\mu\nu} 
+ \al\vp^{\al-1}\na_{\mu\nu}\vp 
+\frac[\al/m-2]\vp^{\al-1}\Dal\vp g_{\mu\nu} 
+\(\al(\al-1)\vp^{\al-2} + \frac[\om/\vp^\be]\) \na_\mu\vp\na_\nu \vp +\cr
&\qquad\qquad
	-\(\al(\al-1)\vp^{\al-2} + \frac[\om/2\vp^\be]\)\na_\la\vp\na^\la\vp g_{\mu\nu}
+\frac[1/2] U g_{\mu\nu}\cr
&
\( \frac[2\om/\vp^{\be-1}] +\frac[2\al(m-1)/m-2]\vp^{\al-1}\)\Dal\vp 
+\(\frac[2\al(\al-1)(m-1)/m-2]\vp^{\al-2} +\frac[\al-\be/\al]\frac[\om/\vp^\be] \)\na_\la\vp\na^\la\vp 
+\frac[1/\al]\vp U' -\frac[m/2]U=T	\cr
&E_i=0\cr
}
\fn$$

If one sets $\be:=2-\al$ and $\om:=-\frac[\al^2(m-1)/m-2]$ these become
$$
\cases{
&\vp^\al R_{\mu\nu} = T_{\mu\nu}  -\frac[1/2]Tg_{\mu\nu} 
+ \al\vp^{\al-1}\na_{\mu\nu}\vp 
+\frac[\al/m-2]\vp^{\al-1}\Dal\vp g_{\mu\nu} 
+\al\frac[2-m-\al/m-2] \na_\mu\vp\na_\nu \vp +\cr
&\qquad\qquad
	+\frac[\al(\al-1)/m-2]\na_\la\vp\na^\la\vp g_{\mu\nu}
-\frac[m-2/4] U g_{\mu\nu}\cr
&
\frac[1/\al]\vp U' -\frac[m/2]U=T	\cr
&E_i=0\cr
}
\fl{BDEQ}$$

In view of the relation between Ricci tensors of conformal metrics, namely
$$
\tilde R_{\al\be}= R_{\al\be} -\frac[m-2/2\vp]\na_{\al\be} \vp-\frac[1/2\vp]\Dal\vp g_{\al\be} + \frac[3m-6/4\vp^2] \na_\al\vp \na_\be\vp
-\frac[m-4/4\vp^2] \na_\ep\vp\na^\ep\vp g_{\al\be}
\fn$$
one can recast field equations \ShowLabel{FEQ1} for Palatini-$f(\calR)$ theories in the equivalent form
$$
\eqalign{
\vp^\al R_{\mu\nu} =&  \vp^{\al-\frac[m-2/2]}\(T_{\mu\nu}-\frac[1/m]Tg_{\mu\nu} \)
+ \frac[m-2/2]\vp^{\al-1}\na_{\mu\nu} \vp
 - \frac[3m-6/4\vp] \vp^{\al-2}\na_\mu\vp \na_\nu\vp+\cr
&+\frac[1/2]\vp^{\al-1}\Dal\vp g_{\mu\nu}
+\frac[m-4/4]\vp^{\al-2} \na_\la\vp\na^\la\vp g_{\mu\nu}
+\frac[1/m] \vp^\al \calR g_{\mu\nu}
}
\fn$$
which is equal to the first equation in \ShowLabel{BDEQ} provided one sets $\al:=\frac[m-2/2]$ and $U:=f(r(\vp))-\frac[2/m-2] \vp^\al r(\vp)$.
Moreover, by these choices the second equation in \ShowLabel{BDEQ} becomes
$$
\frac[1/\al]\vp U' -\frac[m/2]U= \vp^\al r(\vp) -\frac[m/2] f=T
\fn$$
which is equivalent to the master equation in Palatini-$f(\calR)$ theory.

Thus Brans-Dicke theory (with $\al=\frac[m-2/2]$, $\be=\frac[6-m/2]$, and $\om=-\frac[(m-2)(m-1)/4]$ and potential $U= f(r(\vp))-\frac[2/m-2] \vp^\al r(\vp)$) is dynamically equivalent to Palatini-$f(\calR)$ theory.
By {\it dynamically equivalence} we mean there is a one-to-one correspondence between solutions of the two theories
$$
(g_{\mu\nu}, \tilde\Ga^\al_{\be\mu}, \psi^i) 
\quad\longleftrightarrow\quad
(g_{\mu\nu}, \vp, \psi^i)
\fn$$
defined by $\vp=\(f'(\calR)\)^{\frac[2/m-2]}$ (and its inverse defined by $\tilde \Ga=\{\vp g\}$).

Let us stress that this equivalence is established at the level of field equations and solutions. 
We shall establish it at the level of actions below.

Moreover, the two conformal transformations performed in Palatini-$f(\calR)$ theories  are different. In the first place, to solve Palatini-$f(\calR)$ theories, we defined the conformal metric $\tilde g= \vp g$ (leaving the connection unchanged). Then the connection is eliminated by solving the second field equation. Then one goes back to the original metric by the transformation $g=\vp^{-1} \tilde g$. However, at this point the connection $\tilde \Ga$ has been eliminated as an independent field and substituted by $\tilde \Ga=\{\tilde g\}$. Accordingly, the transformation  $g=\vp^{-1} \tilde g$ does act both on the metric and connection (indeed one transforms the whole Ricci tensor $\tilde R_{\al\be}$). In other words these two transformations are not the inverse to each other and in fact one does not obtain the original theory but Brans-Dicke theory which is a purely metric theory.
In the following Section we shall consider the same transformations at the level of Lagrangians to clarify their roles.

\NewSection{Helmholtz Lagrangian for Palatini-$f(\calR)$ Theories }

A first step of transformation at the level of Lagrangian was considered, in dimension $m=4$, in \ref{Magnano}. Let us start by reviewing it in a generic dimension.

For any Palatini-$f(\calR)$ theory one can define the function $\calR=r(\vp)$ by inverting the equation $\vp=\(f'(\calR)\)^{1/\al}$, where we set $\al:=\frac[m-2/2]$. Then one can consider the Helmholtz Lagrangian
$$
L_H= \[\sqrt{g}\(\vp^\al (\calR -r(\vp)) + f(r(\vp))\) + \calL_m(g, \psi)  \]d\si
\fn$$
which is considered as a Lagrangian for the independent fields $(g_{\mu\nu}, \tilde \Ga^\al_{\be\mu}, \vp, \psi^i)$.

Field equations of this Lagrangian are
$$
\cases{
&\vp^\al \tilde R_{(\mu\nu)} = T_{\mu\nu} +\frac[1/2] f(r(\vp)) g_{\mu\nu}\cr
&\tilde \na_\la \( \sqrt{g} \vp^\al g^{\mu\nu}\)=0\cr
&\calR=r(\vp)\cr
&E_i=0\cr
}
\fl{HFEQ}$$

These equations are equivalent to field equations of the original Palatini-$f(\calR)$ theory, namely \ShowLabel{fREQ}
together with the definition of the conformal factor $\vp^\al=f'(\calR)$.
The third equation $\calR=r(\vp)$ is equivalent to $\vp^\al=f'(\calR)$, i.e.~the definition of the conformal factor.
In view of this definition, then the second field equation $\tilde \na_\la \( \sqrt{g} \vp^\al g^{\mu\nu}\)=0$ becomes equal to the second field equations in 
\ShowLabel{fREQ}. The first equations in \ShowLabel{fREQ}. and \ShowLabel{HFEQ}  also becomes equal. Finally, the matter field equations $E_i=0$ are equal in both theories as well.

Accordingly, the correspondence
$$
(g_{\mu\nu}, \tilde\Ga^\al_{\be\mu}, \psi^i) 
\quad\longleftrightarrow\quad
(g_{\mu\nu}, \tilde\Ga^\al_{\be\mu}, \vp, \psi^i)
\fn$$
defined by $\vp= \(f'(\calR)\)^{1/\al}$
does in fact send solutions into solutions.

The Lagrangian $L_H$ is defined on a suitable jet prolongation of the configuration bundle
$$
\calB=\Lor(M)\times_M (M\times \R) \times_M \Con(M)\times_M B
\fn$$
which has coordinates $(x^\mu, g_{\mu\nu}, \vp, \tilde\Ga^\al_{\be\mu}, \psi^i)$.
Let us refer to this choice of independent fields as the {\it Helmholtz frame}.

The Lagrangian $L_H$ is obtained by introducing in Palatini-$f(\calR)$ theory the  momentum $\vp$ conjugated to $\calR$ as an independent variable.

\NewSubSection{Einstein frame}
The Einstein frame (i.e.~the choice of independent fields, or the choice of the representative of the conformal gauge) is a field coordinate system in which the theory seems standard GR.
On the bundle $\calB$ one has coordinates $(x^\mu, g_{\mu\nu},\vp, \tilde \Ga^\al_{\be},  \psi^i)$. However, one can use equivalent fields
$$
(x^\mu, \tilde g_{\mu\nu}= \vp g_{\mu\nu}, \tilde \Ga^\al_{\be}, \vp, \psi^i)
\fn$$
This is simply a field transformation and, accordingly, it does not change the theory. The Lagrangian in the new coordinates reads as
$$
L_E(\tilde g_{\mu\nu}, \tilde \Ga^\al_{\be}, \vp, \psi)= \[\sqrt{\tilde g}\( \tilde g^{\mu\nu} \tilde R_{\mu\nu}+ \vp^{-\frac[m/2]}\(  f\(r(\vp)\)- \vp^{\al}r(\vp)\) \) + \calL_m(\vp^{-1} \tilde g, \psi)  \]d\si
\fn$$
which is in fact a standard Palatini GR Lagrangian with an additional matter field $\vp$ (which enters at order zero in the dynamics) 
and the effective matter Lagrangian
$$
\tilde \calL_m(\tilde g, \vp, \psi)=\calL_m(\vp^{-1} \tilde g, \psi)+ \sqrt{\tilde g} \vp^{-\frac[m/2]}\(  f\(r(\vp)\)- \vp^{\frac[m-2/2]}r(\vp)\)
\fn$$
Since the conformal factor $\vp$ enters at zero order its field equation is algebraic and it inherits its dynamics from the gravitational field mediating together with $\tilde g$ the interaction with matter.  

In this theory, keeping the EPS interpretation of Palatini-$f(\calR)$ theories as ETG, one has dark sources described by the interactions mediated through $\vp$ (together with $\vp$ itself), the free fall of test particles is disctated by $\tilde g$ while operational definitions for distances and clocks (original encoded by the metric $g$) are now described by $g=\vp^{\frac[m-2/2]} \tilde g$.

\NewSubSection{Brans-Dicke frame}

Instead aiming to write everything in terms of $\tilde g$ one can again start from the Helmholtz Lagrangian and do the transformation
$$
\Ga^\al_{\be\mu} = \tilde \Ga^\al_{\be\mu} + \frac[1/2]\(  g^{\al\ep}  g_{\be\mu} -2\de^\al_{(\be} \de^\ep_{\mu)}\)\na_\ep\ln(\vp)
\fl{BDTransf}$$
this time choosing independent fields as  $(g_{\mu\nu}, \Ga^\al_{\be\mu}, \vp, \psi^i)$.
 The transformation has been chosen so that it transforms the independent connection $\tilde \Ga$ as it would change under conformal transformation of the metric if it were the Levi-Civita connection of the corresponding metric.
We shall see below that then field equations in fact will fix $\Ga^{\al}_{\be\mu}$ to be the Levi-Civita connection of the metric $g$.

This transformation needs to be treated more carefully than the one to Einstein frame since the new fields do not depend on old fields only but also on their derivatives (on the first derivatives of $\vp$ in particular).
Let us define the bundle $\calC=\Lor(M)\times_M (M\times \R) \times_M \Con(M)\times_M B $
 with fibered coordinates $(x^\mu, g_{\mu\nu}, \vp, \tilde \Ga^\al_{\be\mu}, \psi^i)$ for the Helmholtz frame, $(x^\mu, \tilde g_{\mu\nu}, \vp, \tilde \Ga^\al_{\be\mu}, \psi^i)$ for the Einstein frame, $(x^\mu, g_{\mu\nu}, \vp,  \Ga^\al_{\be\mu}, \psi^i)$ for the Brans-Dicke frame.
Let us also define the bundle $\calB=\Lor(M)\times_M J^1(M\times \R) \times_M \Con(M)\times_M B$
so that transformation $\ShowLabel{BDTransf}$ induces a map 
$$
\Phi: \calB\arr\calC
\fn$$
The original Helmholtz Lagrangian is a horizontal form on $J^1\calC$ and the bundle $J^1\calC$ can be pulled-back along the transformation 
$\Phi$ on $\calB$, as described by the following diagram
$$
\begindc{\commdiag}[1]
\obj(180,130)[JC]{$J^1\calC$}
\obj(110,150)[JB]{$\Phi^\ast J^1\calC$}
\obj(110,100)[B]{$\calB$}
\obj(180,80)[C]{$\calC$}
\obj(110,30)[M2]{$M$}
\obj(180,30)[M3]{$M$}
\mor{B}{M2}{}
\mor{C}{M3}{}
\mor{JB}{JC}{$J\Phi$}
\mor{JB}{B}{}
\mor{JC}{C}{$\pi^1_0$}
\mor{B}{C}{$\Phi$}
\mor{M2}{M3}{}[\atleft, \solidline] \mor(110,33)(180,33){}[\atleft, \solidline]
\enddc
\fn$$

The Helmholtz Lagrangian $L_H$ can be pulled-back on the bundle
$$
\Phi^\ast J^1\calC = J^1\Lor(M)\times_M J^2(M\times \R)\times_M J^1 \Con(M)\times_M J^1B
\fn$$
along the map $J\Phi$  to obtain the Lagrangian
$$
L^\ast (j^1g, j^2\vp, j^1\Ga, j^1\psi)= L_H(j^1g, j^0\vp, j^1\tilde \Ga(\Ga,g, j^1\vp) , j^1\psi) 
\fn$$
Since we know the variation of $L_H$ to be in the form
$$
\de L_H= E_{\al\be} \de g^{\al\be} + E \de \phi + E_\la^{\mu\nu} \de \tilde \Ga^\la_{\mu\nu} + E_i\de\phi^i + (\Div)
\fn$$
then the variation of the Lagrangian $L^\ast$ is 
$$
\eqalign{
\de L^\ast =& E_{\al\be} \de g^{\al\be} + E \de \phi + E_\la^{\mu\nu} \de \tilde \Ga^\la_{\mu\nu} + E_i\de\phi^i + (\Div)=\cr
=&E_{\al\be} \de g^{\al\be} + E \de \phi + E_\la^{\mu\nu}\(\de \Ga^\la_{\mu\nu} 
+ \Frac[\del \tilde\Ga^\la_{\mu\nu}/\del g^{\al\be}]\de g^{\al\be} +\Frac[\del \tilde\Ga^\la_{\mu\nu}/\del \vp]\de \vp
+\Frac[\del \tilde\Ga^\la_{\mu\nu}/\del \vp_\si]\na_\si \de \vp\) + E_i\de\phi^i + (\Div)=\cr
=&\(E_{\al\be}+ E_\la^{\mu\nu} \Frac[\del \tilde\Ga^\la_{\mu\nu}/\del g^{\al\be}]\)  \de g^{\al\be} 
+ \(E + E_\la^{\mu\nu} \Frac[\del \tilde\Ga^\la_{\mu\nu}/\del \vp] 
 - \na_\si \(E_\la^{\mu\nu}\Frac[\del \tilde\Ga^\la_{\mu\nu}/\del \vp_\si]\)\) \de \phi + E_\la^{\mu\nu}\de  \Ga^\la_{\mu\nu} + E_i\de\phi^i + (\Div)\cr
}
\fn$$
Accordingly field equations for the Lagrangian $L^\ast$ are
$$
\cases{
&E_{\al\be}+ E_\la^{\mu\nu} \Frac[\del \tilde\Ga^\la_{\mu\nu}/\del g^{\al\be}]=0\cr
&E + E_\la^{\mu\nu} \Frac[\del \tilde\Ga^\la_{\mu\nu}/\del \vp]  - \na_\si \(E_\la^{\mu\nu}\Frac[\del \tilde\Ga^\la_{\mu\nu}/\del \vp_\si]\)=0\cr
&E_\la^{\mu\nu}=0\cr
&E_i=0\cr
}
\fn$$
which though different are equivalent to the ones of the Helmholtz Lagrangian, and hence equivalent to the original Palatini-$f(\calR)$ theory.

In fact, matter and connection equations are always unchanged just written in the new variables.
Then in particular the field equation of the connection keep prescribing that $\tilde \Ga=\{\tilde g\}$, i.e., equivalently, that $\Ga=\{g\}$.
Once this equation is satisfied, then the additional terms in the field equations of the metric $g$ and conformal factor $\vp$ vanishes and also these equations determines the same conformal factor and metric as the Helmholtz Lagrangian.
In conclusion also the Lagrangian $L^\ast$ is dynamically equivalent to the Helmholtz Lagrangian and hence to Brans-Dicke 
and the original Palatini-$f(\calR)$ theory.

Let us stress that also we showed that $L^\ast$ is an equivalent formulation of the metric Brans-Dicke theory in a metric-affine form.
If one takes the Brans-Dicke theory seriuosly though, the metric $g$ describes light cones, free fall of test particles and observational protocols of distances.
If one keeps stuck to the EPS interpretation of the original Palatini-$f(\calR)$ theory, then $g$ describes light cones and observational protocols, while free fall of test particles is described by the metric $\tilde g= \vp \cdot g$.

\NewSection{Conclusions and Perspectives}

Palatini-$f(\calR)$ are not the most general ETG and are relatively disregarded in the literature for a number of issues which we would like to summarize and discuss hereafter. 

\NewSubSection{Solar System tests}
The first issue is that solar system tests rule out Brans-Dicke theories (with no potentials) or at least put strong constraint on the parameter $\om$.
In dimension $m=4$ we obtained the value $\om=-\frac[3/2]$ which is almost certainly ruled out.
Although a detailed discussion would be needed to deal with potentials, we have noticed that in view of the analysis and EPS interpretation Brans-Dicke theories are dynamically equivalent to Palatini-$f(\calR)$ though one has different free fall in the two cases. 
While in the classical Brans-Dicke theory free fall is described by $g$, in Palatini-$f(\calR)$ it is described by $\tilde g$. Being the two metrics conformal they share the same light cones and null geodesics. 
However, they have  different timelike geodesics. 
For example, if one assumes Mercury to follow a  geodesic orbit with respect to $\tilde g$ there are plenty of $f(\calR)$ models (all of them corresponding to a Brans-Dicke with $\om=-\frac[3/2]$) which pass the classical precession tests; see \ref{Kepler}, \ref{TM1}, \ref{TM2}.  

As a matter of fact, free fall of test particles is an independent assumption in field theory which is not determined by the variational framework which instead fixes the interactions between fields. 
Of course, one can obtain equations for curves from field equations (by means of characteristics, eikonal approximation, motion of wave packets, solutions supported on curves). However, none of these methods can prescribe an interpretation. For example, let us consider particles associated to a Klein-Gordon field $\phi$.
Klein-Gordon field equations read as
$$
(\Dal+m^2)\phi=0
\fn$$ 
and one can associate to these equations geodesics of the metric used to define the box operator (e.g.~through characteristics).
Nicely this independent of the mass $m^2$ in the Klein-Gordon operator.

Can one say that there is a test particle falling along $g$-geodesics?

It mostly depends if we believe that the field $\phi$ is associated to test particles. In fact one can simply perform a conformal transformation
in the Klein-Gordon sector and define a new field $\tilde \phi= \vp^{-\frac[1/2]} \phi$ which obeys a new Klein-Gordon field equation 
$$
\( \tilde{\Dal} + \tilde m^2\) \tilde \phi=0
\fn$$ 
to which one associates $\tilde g$-geodesics; see \ref{F1},  \ref{F3}, \ref{TM1}.
Whether tests particles are described by $\phi$ or $\tilde \phi$ is again a matter of definition.

If one believes in this freedom, then new possibilities arise. For example, one could ask whether it is sound a model in which quantum theory for the particle associated to the Klein-Gordon field is developed in a reference frame in which $g$ is locally almost Minkowskian (i.e.~by using $\phi$), while free fall is dictated by $\tilde g$ (i.e.~by using $\tilde\phi$).

Of course, it would be strange to do quantum mechanics in a reference frame which is not free falling. However, one can trivially remark that  colliders on the Earth are not free falling; they are {\it approximately} free falling, in the sense that gravitational effects are weak compared to the interactions considered and experiments are short so that one can {\it neglect} gravity.
Let us stress that in most models (e.g.~in cosmology) it is reasonable to expect the differences between $g$ and $\tilde g$ to become manifest on very large scales (including very long time intervals) compared to the scales of the whole universe.
Modifications due to conformal factor can be locally neglected compared to gravitational effects which can in turn be neglected compared to other interaction studies in colliders.

Then all one should do is trying to define such a model and make predictions that can be falsified by observations.
For example falsification can come from baryogenesis or structures formation. It appears very unlikely that they come from particle physics in colliders or quantum effects in standard models.

\NewSubSection{Matter spectra}

We can see stars very distant from us and observe  absorption lines of the gas around them very precisely.
Since absorption lines are determined by physical constants (speed of light, Planck constant) and coupling constants (mass and charge of the electron).
There are very tight constraints about possible changes of these constants also over (most of) the life of the universe.

From the analysis we did above, one can see that matter equations do not change while the frame is changed. 
The coupling constants among matter and matter (e.g.~the electric charges) are not affected by conformal transformations.
The conformal factor just affect the coupling between matter and gravity.
This effect can be seen as dark sources, or used to redefine the Newton universal constant to define an effective coupling constant $G_{eff}$ which of course depends on the point through the conformal factor.

Constraints on $G$ are considerably less tight than the ones about, e.g., electron charges. Usually one finds that $G$ is assumed to be constant up to $10\%$ over the life of the universe. Tighter constraints may comes from baryogenesis and structure formations which though strongly depend on the
specific models and need to be investigated in details.

In any event, in cosmological applications one expects a conformal factor which depends just on the age of the universe and the function $f$ is chosen 
so that it does not violate obvious constraints (e.g.~producing in vacuum a tiny positive cosmological constant); see \ref{Olmo}.  
There are a huge family of functions $f$ to be chosen so that the effective coupling constant $G_{eff}$ varies very slowly in time and space so that one can assume it to be constant over the Solar System over centuries. Now it is not hard to get that we measured $G$ in the solar system in the last few centuries
so that for us $G\simeq G_{eff}$, i.e.~it corresponds to a conformal factor $\vp\sim 1$ of the order of unity.
This is not a coincidence, but the consequence of our conventions on physical units which are in turn determined by experiments done here and now.

On the other hand, we often observe systems far away in space and back in time, often at distances comparable with the  dimension of the whole universe.
However, conformal factor can be conjectured to be constant during the phenomena we observe. 
As a matter of fact we did not make observations or experiments that are sensitive to changes of conformal factors.
For example, we did not measure $G$ over billion of years.

\NewSubSection{No-go theorem for polytropic stars}

It has been noticed that in Palatini-$f(\calR)$ theories if one matches a star model assuming polytropic equation of state (EoS), there are 
physical values of the polytropic parameters (e.g.~the values used to model neutron stars) for which the inner solution develop an essential singularity near the surface of the star; see \ref{Sout}. This has been used to claim that generic  Palatini-$f(\calR)$ theories are unphysical since they do not allow simple models of stars. 

After that it was argued that the result strongly depends on EoS which are by their nature an approximation.
The singularity develop at pressure unphysically low, so that any slight changes of the matter sector may prevent the singularity formation; see \ref{Olmo-No-go}.

We recently analyzed the result and show further weakness in its interpretation; see \ref{Mana}.
In the first place the result mathematically relies on regularity of the function $f(\calR)$. If continuous but non-smooth functions $f(\calR)$ are allowed
then one can define models in which the previously diverging models are now regular.

Moreover and more importantly, the singularity is developed when matching the metric $g$. One can show that matching $\tilde g$ no singularity is formed
(of course since $\tilde g$ obeys standard GR though with extra dark sources).
That means that if $\tilde g$ matches when $g$ does not, the singularity comes from the conformal factor.
In fact one can explicitly show that the the conformal factor is continuous but not differentiable at the surface.
Solutions with $\calC^1$ matching usually needs extra source shells at the surface which in view of what noticed in \ref{Olmo-No-go} can call for a modification of EoS at the surface preventing the formation of the singularity.

And in any event, in view of EPS interpretation, one should match $\tilde g$ not $g$ and in Einstein frame there is no singularity.
Also when the matching of $g$ is regular, matching $g$ or $\tilde g$ can be shown to be different and one expects that also 
in that case the physical model of a star should deal with matching $\tilde g$.
And matching $\tilde g$ is not more difficult than in standard GR.

\NewSubSection{Standard clocks}

Perlick considered a definition of standard clock which is viable in Weyl geometries and consequently in Palatini-$f(\calR)$ theories; see \ref{Perlick}.
Basically he defined a standard clock as a clock for which covariant acceleration is normal to covariant velocity. This extends the request of being free falling which does not apply to clocks on Earth in the first place.
He also proved that any clock can be made standard by properly changing parametrization.
Also one can prove that any congruence of clocks can be made standard by selecting a suitable representative of the conformal class; see \ref{Polistina}.
Moreover, any congruence of clock does {\it select} (i.e.~uniquely determine) a representative of the conformal class for which they are standard clocks.

Hence one could ask what happen if the atomic clocks happened not to be standard for the metric governing the free fall of test particles?
That they would determine a new representative of the conformal factor quite naturally.

Then the issue becomes now if could we experimentally test whether the atomic clocks are standard for $\tilde g$ or not.
Once again one should assume they are not, make predictions and test.
  
\NewSubSection{Distances measurements}

Chandrasekhar  limit is determined by a balance of quantum effects and gravitational attraction.
If in a Palatini-$f(\calR)$ theory the Newton constant acts effectively through $G_{eff}$ and depends on time, while quantum mechanics is not depending on the conformal factor (otherwise that would affect emission lines) then the limit would itself depend on time.
That means that far away supernovae Ia may have a different absolute luminosity since they could be fired at a different critical mass.
And one has to stress that supernova Ia are used as standard candels just because they are assumed the have the same absolute luminosity.

Could we experimentally see if supernovae Ia were not standard?

\ms

All the issues suggest the following model interpretation for Palatini-$f(\calR)$ theories.
Visible standard baryonic matter is described by the matter fields $\psi$ which couples to $g$ in $\calL_m$.
According to EPS the free fall is described by $\tilde g$ and quantum mechanics is described in the frame of $g$.
Quite naturally atomic clocks are not necessarily standard for $\tilde g$ but for $g$, which implies that distances and time intervals are described by $g$.
Due to this different frame between gravitational and quantum physics visible matter acts as gravitational source together with dark sources appearing effectively in the Einstein frame, but disappearing at fundamental level in the original Jordan frame.

Can we detect experimentally a possible shift of frame between gravitational and fundamental quantum physics?

This conjectured shift is not tiny at the cosmological scales and it requires the tuning of many different effects. Just for that it should not be difficult to be disproven.
Let us finally stress that this proposal is simply more general that standard GR. It only by assuming more general frameworks that one can conceive tests that can confirm or disprove the old ones. Even if finally standard GR will be confirmed and ETG disproven on the basis of observations by doing that we shall find a better and more solid foundation for interpretation of gravitational theories.

\Acknowledgements

We acknowledge the contribution of INFN (Iniziativa Specifica QGSKY), 
the local research project {\it Metodi Geometrici in Fisica Matematica e Applicazioni} (2013) of Dipartimento di Matematica of University of Torino (Italy). 
This paper is also supported by INdAM-GNFM.

We are grateful to M.Ferraris and S.Capozziello for discussions and comments.
We also thanks T.Sotiriou and O.Bertolami for interesting comments.

This work is dedicated to the memory of Mauro Francaviglia. 
It stems from a problem he gave in a Ph.D. course to one of the authors (S.G.).

\ShowBiblio

\end